\documentclass[pra,longbibliography,   amsmath, amssymb, 
  floatfix]{revtex4-2}
\usepackage{amsmath,graphics, graphicx, bm, natbib, color}
\usepackage[capitalise]{cleveref}
\usepackage{graphics, graphicx, bm, natbib, color, xcolor}
\usepackage{subcaption}
\usepackage{comment}
\usepackage{appendix}

\newcommand{\rem}[1]{}

\newcommand{\ud}{\mathrm{d}}
\newcommand{\diag}{\text{diag}}
\newcommand*{\vertbar}{\rule[-1ex]{0.5pt}{2.5ex}}

\newenvironment{psmallmatrix}
  {\left(\begin{smallmatrix}}
  {\end{smallmatrix}\right)}

\begin{document}

\author{Wijnand Broer}

\affiliation{Institute of Physics, University of Chinese Academy of Sciences, 100190 Beijing, China}

\author{Bing-Sui Lu}
\affiliation{Division of Physics and Applied Physics, Nanyang Technological 
University, Singapore 637371, Singapore}

\author{Rudolf Podgornik}
\affiliation{School of Physical Sciences, University of Chinese Academy of Sciences, Beijing 100049, China}
\affiliation{Institute of Physics, Chinese Academy of Sciences, 100190 Beijing, China }
\affiliation{Kavli Institute for Theoretical Sciences, University of Chinese Academy of Sciences, Beijing 100049, China}
\affiliation{Wenzhou Institute of the University of Chinese Academy of Sciences, Wenzhou, Zhejiang 325000, China}
\altaffiliation{Department of theoretical physics, J. Stefan Institute, , 1000 Ljubljana, Slovenia and Department of Physics, Faculty of Mathematics and Physics, University of Ljubljana, 1000 Ljubljana, Slovenia}
 \email{podgornikrudolf@ucas.ac.cn}

\title{{Qualitative   chirality effects on the Casimir-Lifshitz torque with liquid crystals}}

\date{\today}

\begin{abstract}

{ We model a cholesteric liquid crystal as a planar uniaxial multilayer system, where the orientation of each layer differs slightly from that of the adjacent one. This allows us to analytically simplify the otherwise acutely complicated calculation of the Casimir-Lifshitz torque. Numerical results differ appreciably from the case of nematic liquid crystals, which can be treated like bloc birefringent media. In particular, we find that the torque deviates considerably from its usual sinusoidal behavior as a function of the misalignment angle. In the case of a birefringent crystal faced with a cholesteric liquid one, the Casimir-Lifshitz torque decreases more slowly as a function of distance than in the nematic case. In the case of two cholesteric liquid crystals, either in the homochiral or in the heterochiral configuration, the angular dependence changes qualitatively as a function of distance. In all considered cases, finite pitch length effects are most pronounced at distances of about 10 nm.  }
\end{abstract}
\maketitle

\section{Introduction}

Casimir-Lifshitz interactions \cite{Casimir48,Lifshitz55, Lifshitz61} are macroscopic dispersion forces that arise from quantum mechanical and thermal fluctuations of the electromagnetic field. {The name `dispersion forces'  originates from the fact that their properties are governed by the dielectric and magnetic dispersion of the interacting materials \cite{vdWBook,CasimirBook, BuhmannBookI, Woods2016Review}.} Casimir-Lifshitz forces share the same physical origin as the van der Waals-London forces, which conventionally denote the interaction in cases where retardation effects are negligible. 

{These forces are studied for both fundamental and practical reasons. From a fundamental point of view, the Casimir-Lifshitz force plays an important role in the search for deviations from Newtonian gravitation in short range gravitation experiments. Its study can facilitate the precise comparisons between theoretical predictions and experimental data \cite{Sedmik2020}.  More practically, Casimir-Lifshitz interactions affect the actuation dynamics of nano- and micro-mechanical systems, such as switches, cantilevers, and actuators at a sub-micrometer length scale \cite{Serry1995,Broer2013, Broer2015, Mehdi2015, Tajik2018PRE, Munday2021Review}. The Casimir-Lifshitz interactions also underpin the stability of colloidal and biophysical macromolecular systems. These forces represent one of the pillars of the fundamental Deryaguin-Landau-Verwey-Overbeek theory of colloid stability \cite{Israelachvili}. } 

{Precise evaluation of the Casimir-Lifshitz force between real materials requires a detailed knowledge of the materials' electromagnetic susceptibilities as inputs \cite{Lifshitz55, Lifshitz61}. Because this interaction is mediated by virtual photons, the frequency of which cannot be controlled directly, the Casimir-Lifshitz interaction is a broadband phenomenon. Consequently, the electromagnetic susceptibilities of the materials involved must be known - either theoretically or experimentally - as a function of frequency in a sufficiently broad range. 
Another requirement to evaluate the Casimir-Lifshitz interaction  is solving Maxwell's equations for the given geometric configuration of the interacting bodies. {This latter aspect of the Casimir-Lifshitz calculations is the focus of this paper.} } 

{The presence of anisotropies, either in the susceptibilities or the morphological anisotropy induced by the shape of the interacting bodies \cite{Hopkins2015}, implies the existence of not only forces but also  Casimir-Lifshitz torques between the interacting bodies \cite{Parsegiantorque}. Specifically, the dielectric anisotropy in the plane of reflection, which will exclusively concern us in what follows,  creates a dielectric contrast in the azimuthal direction, which gives rise to the  Casimir-Lifshitz torque attempting to align the orientations of both materials\cite{Ginzburg75}.   An exact analytical description of the Casimir-Lifshitz torque between two planar uniaxially anisotropic half spaces was originally based on the direct solution of the Maxwell equations and the pertaining dispersion equation by Barash more than four decades ago \cite{Barash1978}. An alternative methodology applicable to the Casimir-Lifshitz torque can be based on the solutions of the Maxwell's equations by interpreting them as an eigenvalue problem. This method is known as the transfer matrix method.  It is especially suited for describing electromagnetic field propagation through anisotropic materials \cite{Berreman1972, Yeh1979, Lekner1991}, and can be used to derive the Casimir torque between semi-infinite birefringent plates \cite{Shao2005,Philbin2008}. The results of the transfer matrix methodology were later shown to be consistent with the direct method of Barash \cite{Broer2019}. Some recent examples of the application of this method in the context of Casimir physics include biaxial materials, \cite{Emelianova2020} Weyl semi-metals, \cite{ChenLiang2020, Farias2020} and magnetic ferrite slabs \cite{Zeng2020}.  }

{The first experimental observation of Casimir-Lifshitz torques has been accomplished only recently, by relying on the anisotropic dielectric response of nematic liquid crystals \cite{Somers2018}. It seems that further experiments on Casimir-Lifshitz torques with liquid crystals are within experimental reach presently or in the near future. Apart from the Casimir-Lifshitz torques, liquid crystals exhibit diverse other Casimir fluctuation phenomena  and so might prove to be an abundant source of a variety of distinct Casimir phenomena \cite{Karimi_Pour_Haddadan_2014, Haddadan1}. The simplest way to theoretically describe the anisotropy of liquid crystals is by approximating them as uniaxial half spaces with a unique orientation \cite{Somers2015}. This essentially approximates the liquid crystal as a birefringent half space. In what follows we will refer to this approximation as {\sl 'half space approximation'}. This is indeed a valid approach if the liquid crystal is aligned, such as in the case of nematic liquid crystals \cite{Kornilovitch2012, *Kornilovitch2018}.
However, a cholesteric liquid crystal actually consists of a sequence of many thin, planar, uniaxially oriented layers, whose orientation varies periodically in space with a periodicity given by the cholesteric pitch \cite{Ryabchun2018}. This type of layered system has been shown to be amenable to the transfer matrix method \cite{Berreman1972} as electromagnetic waves in layered anisotropic media can be described by an algebra of 4$\times$4 matrices \cite{Veble2009}.

However, a major complication arises from the fact that the cholesteric liquid crystal consists of a large number of layers ($\sim10^4$) which generally implies a product of an equally large number  of 4$\times$4 matrices. Even though the non-retarded limit reduces the size of the matrices to 2$\times$2, it does not reduce the number of matrix multiplications \cite{Rudi2004,Kornilovitch2012,Bing2016, Bing2018}}. {Since the computational complexity of matrix multiplication increases linearly with the number of multiplications, an explicit analytical expression for the transfer matrix can reduce this complexity by up to four orders of magnitude. Hence here we propose a  way to analytically simplify this problem and derive an explicit expression for the composite transfer matrix for a specific planar multilayer structure that models a cholesteric liquid crystal. We have taken advantage of the facts that each layer is very thin, and that the difference in orientation between each successive layer is small \cite{Ryabchun2018} .} { This allows us to write an analytic result for the transfer matrix based on the Baker-Campbell-Hausdorff (BCH) formula \cite{HighamBook2008} as a second order expansion in this parameter, $\delta$, which is inversely proportional to the pitch length of the cholesteric liquid crystal. The form of the transfer matrix is then obtained as a function of the ordinary and extraordinary eigenvalues of the Maxwell equations, averaged over the orientations in the multilayer system.  {Consequently, our solutions enable us to analyze the interaction between two semi-infinite homochiral or heterochiral layers }and thus quantify macroscopically the effects of chirality. Our results point to a realistic possibility of measuring the effects of chirality in the macroscopic Casimir-Lifshitz torque for interaction geometries involving oriented cholesteric liquid crystals.} Note that this is different from effects of chirality of an intervening medium \cite{Barcellona2017, Jiang2019, Safari2020}.

The paper is organized as follows. After the introduction we will discuss the solutions of the Maxwell equations for a general planar anisotropic multilayer in section \ref{sec:Matrix}.  Next we will introduce the simplification for the cholesteric liquid crystal configuration in section \ref{sec:CLC}, the numerical implementation of which is then covered in section \ref{sec:Results}. Conclusions can be found in section \ref{sec:Conclusions}.

\section{Planar anisotropic multilayer}\label{sec:Matrix}

\subsection{Transfer matrix}

\begin{figure}[!htbp]
 \includegraphics[width=0.9\textwidth]{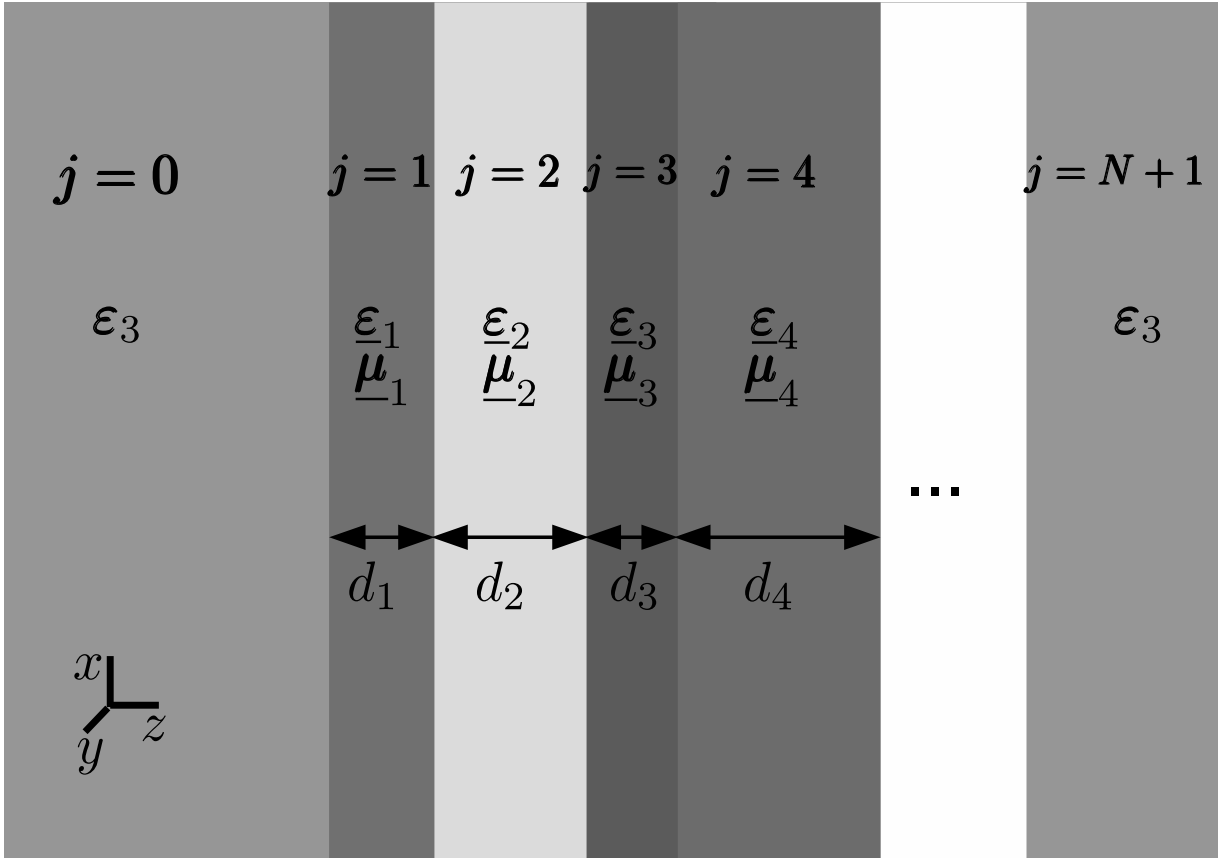}
 \caption{One dimensional stratification of $N$ finite thickness  anisotropic dielectric slabs representing an arbitrary planar multilayer stack. 
 The stratified media, $i = 1,N$, with generally anisotropic layers are embedded in an isotropic bathing medium.  Each layer has an arbitrary material composition, thickness and orientation.  \label{fig:geometry}}
\end{figure}

{Here we will formulate the theory of Casimir-Lifshitz interactions in a system composed of a one dimensional layering of finite thickness  anisotropic dielectric slabs. {(See \cref{fig:geometry})} {Later on, we will use this geometry to model a cholesteric liquid crystal.}  Our approach is based on an application of the transfer matrix method for the electromagnetic field modes propagating through  one-dimensional stratifications.  This method was  introduced as an algebra of 4 $\times$ 4 matrices by Berreman \cite{Berreman1972} and later reformulated by Yeh \cite{Yeh1979}. {Here we will follow the latter formulation, which is based on the change of basis transform in linear algebra. However, both formulations are equivalent \cite{Passler2017}.  }
 We assume that the anisotropic stratification coincides with the direction of the cholesteric axis, oriented along the $z$-direction and perpendicular to the translational $x$-$y$-plane of symmetry of the {laboratory coordinate} system. {(See \cref{fig:geometry})}}

We start by formulating Maxwell's equations as an eigenvalue problem, as we have done before for the case of single interfaces \cite{Broer2019}. Assume that the electromagnetic (EM) wave propagates within one of the anisotropic slabs labeled as $j$ 
whose anisotropic plane is facing the surface, which is defined as the $x$-$y$-plane in the laboratory coordinate system. Furthermore, the magnetic and electric anisotropy axes are assumed to be identical, implying that the electric permittivity and the magnetic permeability are given by:

\begin{equation}\label{eq:eps_tensor}
 \underline{\pmb{\varepsilon}}_j(\omega)=\begin{psmallmatrix}
	\varepsilon_{jx}\cos^2\theta_j+\varepsilon_{jy}\sin^2\theta_j&(\varepsilon_{jx}-\varepsilon_{jy})\sin\theta_j\cos\theta_j&0\\
	(\varepsilon_{jx}-\varepsilon_{jy})\sin\theta_j\cos\theta_j&\varepsilon_{jx}\sin^2\theta_j+\varepsilon_{jy}\cos^2\theta_j&0\\
	0&0&\varepsilon_{jz}
\end{psmallmatrix}
\end{equation}

\begin{equation}\label{eq:mu_tensor}
 \underline{\pmb{\mu}}_j(\omega)=\begin{psmallmatrix}
	\mu_{jx}\cos^2\theta_j+\mu_{jy}\sin^2\theta_j&(\mu_{jx}-\mu_{jy})\sin\theta_j\cos\theta_j&0\\
	(\mu_{jx}-\mu_{jy})\sin\theta_j\cos\theta_j&\mu_{jx}\sin^2\theta_j+\mu_{jy}\cos^2\theta_j&0\\
	0&0&\mu_{jz},
\end{psmallmatrix}
\end{equation}
where $\theta_j$ denotes the angle between the  optic axis of layer $j$  and the $x$-axis of laboratory's coordinate system. It must be stressed that the entries of both $\underline{\pmb{\varepsilon}}_j$ and $\underline{\pmb{\mu}}_j$ depend on frequency, but that the argument will be suppressed from now on. {Because of this dependence it is convenient to represent the electromagnetic fields in the frequency domain as well.  After also transforming to Fourier space, the {vectorial Maxwell} equations are  given by}

\begin{equation}\label{eq:kcross}
\mathbf{k}\times\mathbf{E}=\frac{\omega}{c}\mathbf{B},\qquad \mathbf{k}\times\mathbf{H}=-\frac{\omega}{c}\mathbf{D}
\end{equation}
where $\mathbf{B}=\underline{\pmb{\mu}}\cdot\mathbf{H}$ and $\mathbf{D}=\underline{\pmb{\varepsilon}}\cdot\mathbf{E}$. \cref{eq:kcross} is a system of six linear equations with six unknowns, four of which are independent. {Elimination of the $z-$component of $\mathbf{E}$ and $\mathbf{H}$  leads to the following 4 dimensional eigenvalue equation:  \cite{Rosa2008}}

\begin{equation}\label{eq:3Dk}
  \underline{\pmb{Q}}_j \pmb{\Psi}=q ~\pmb{\Psi},
\end{equation}
for the 4 dimensional EM field vector $\pmb{\Psi} = (E_x, E_y, H_x, H_y)^{T} $.  $\underline{\pmb{Q}}_j$ denotes a 4$\times4$ anti-diagonal block matrix whose explicit entries are given in the supplemental material, section I. 

The four eigenvector solutions of \cref{eq:3Dk} represent the orientation dependent modes, which both can propagate forward and backward. They are denoted by $\pmb{\Psi}_{jo}^{\pm}$ and $\pmb{\Psi}_{je}^\pm$, where the subscripts $e$ and $o$ denote extraordinary and ordinary, respectively (for the case of birefringent media), and the superscripts $+$ and $-$ indicate forward and backward propagating modes, respectively. The mode eigenvalues are given by

\begin{subequations}
\begin{gather}
\label{eq:qjo}q^\pm_{jo}= \pm \sqrt{\varepsilon_{jy}\mu_{jx}\omega^2/c^2-(\mu_{jx}/\mu_{jz})k^2_\rho\cos^2(\theta_j-\eta)-k^2_\rho\sin^2(\theta_j-\eta)}\\
\label{eq:qje}q^\pm_{je}= \pm \sqrt{\varepsilon_{jx}\mu_{jy}\omega^2/c^2-(\varepsilon_{jx}/\varepsilon_{jz})k_\rho^2\cos^2(\theta_j-\eta)-k_\rho^2\sin^2(\theta_j-\eta)}
\end{gather}
\label{eq:eigenvalues}
\end{subequations}
where $k_\rho$ denotes the radial component of the wavevector, and $\eta$ represents the azimuthal component of the EM wave propagation direction.



Here we formulate a 4$\times$4 transfer matrix formalism for a one-dimensional stratified  system composed planar slabs of finite thickness with in-plane anisotropy (See Fig. \ref{fig:geometry}),  based on the methodology of Ref.\cite{Yeh1979}.

The matrix in \cref{eq:3Dk} has distinct eigenvalues so it is diagonalizable. This brings us to the so called \emph{propagation matrix}

\begin{equation}
\underline{\pmb{P}}_j\equiv\mathrm{diag}(\exp(-i q_{je} d_j), \exp(i q_{je} d_j), \exp(-i q_{jo} d_j), \exp(i q_{jo} d_j)).
\label{eq:propmat}
\end{equation}
Multiplication from the left of the matrix $\underline{\pmb{P}}_j$ by the EM field vector represents propagation within an anisotropic slab in the $z$-direction over a distance $d_j$, the thickness of the $j$th layer:  
\[ \pmb{\Psi}(z+d_j)=\underline{\pmb{P}}_j\cdot\pmb{\Psi}(z),					\]
where $\pmb{\Psi}$ again denotes a 4 dimensional EM wave vector. $\underline{\pmb{P}}_j$ is a valid representation only in the eigenvector basis where the matrix is diagonal. Hence it is transformed to the  $xy$ basis as follows

\begin{equation}\label{eq:ChangeOfBasis}
\exp\left({-i\underline{\pmb{Q}}_{j}d_j}\right)= \underline{\pmb{S}}_j\cdot\underline{\pmb{P}}_j\cdot\underline{\pmb{S}}_j^{-1},
\end{equation}
where the columns of $\underline{\pmb{S}}_j$ consist of the eigenvectors of Maxwell's equations: 

\begin{equation}
                     \underline{\pmb{S}}_j=  \begin{pmatrix}
                              \vertbar & \vertbar & \vertbar & \vertbar\\
                              \pmb{\Psi}_{je}^+ & \pmb{\Psi}_{je}^- & \pmb{\Psi}_{jo}^+ & \pmb{\Psi}_{jo}^-\\
                              \vertbar & \vertbar & \vertbar & \vertbar
                             \end{pmatrix}
\label{eq:SDef}
\end{equation}
which is analogous to what is called `rotation matrix' in Ref. \cite{Veble2009}. In \cref{eq:SDef} the normalization constants of the eigenvectors can be omitted because $\underline{\pmb{P}}_j$ is multiplied by the inverse of $\underline{\pmb{S}}_j$ from the right. The total transfer matrix that represents the EM wave propagation through all $N$ layers is given by

\begin{equation}
\underline{\pmb{T}}_N\equiv\mathrm{...} \underline{\pmb{S}}_{j-1}\cdot\underline{\pmb{P}}_{j-1}\cdot\underline{\pmb{S}}_{j-1}^{-1}\cdot \underline{\pmb{S}}_j\cdot\underline{\pmb{P}}_j\cdot\underline{\pmb{S}}_j^{-1}\cdot \underline{\pmb{S}}_{j+1}\cdot\underline{\pmb{P}}_{j+1}\cdot\underline{\pmb{S}}_{j+1}^{-1}\mathrm{...}
\equiv \prod_{j=1}^{N}\underline{\pmb{S}}_j\cdot\underline{\pmb{P}}_j\cdot\underline{\pmb{S}}_j^{-1}
\label{eq:TN}
\end{equation}
where the product symbol denotes matrix multiplication from the right for each subsequent index.

Next we have to consider what happens outside the slab, i.e. the cases $j=0$ and $j=N+1$. The EM field propagating through the isotropic embedding medium can be written as a linear combination of the $s$- and $p$-polarized modes.  By the same token as before, since the $s$- and $p$-polarized modes are the eigenmodes for  isotropic media, a matrix can be constructed with the $s$- and $p$-polarized mode eigenvectors as columns for $j=0$:

\begin{equation}
\underline{\pmb{S}}_0=\begin{pmatrix}
                              \vertbar & \vertbar & \vertbar & \vertbar\\
                              \pmb{\Psi}_{js}^+ & \pmb{\Psi}_{js}^- & \pmb{\Psi}_{jp}^+ & \pmb{\Psi}_{jp}^-\\
                              \vertbar & \vertbar & \vertbar & \vertbar
                             \end{pmatrix}
\label{eq:S0}
\end{equation}
the explicit form of which can be found in the supplemental material, section I. However, the case $j=N+1$ is a little different. {Here only forward propagating modes exist, which does not affect} {the matrix itself but instead only affects the vector on which the matrix acts}. {After all, in the $s-$ and $p-$mode basis, this information can be represented by two nonzero components, which represent the amplitudes of the forward propagating $s-$ and $p-$polarized modes, (see Section \ref{sec:Fresnel} for details).} Note that the matrices of \cref{eq:SDef,eq:S0} are singular at zero frequency. However, physically 'zero frequency' corresponds to a static limit where the frequency is small on the scale of the main absorption frequency of the materials. Hence within this limit it is possible to obtain the inverses of the matrices of \cref{eq:SDef,eq:S0}. {See also supplemental material, section V for more details on how to take the non-retarded limit}.

%
%
Since \cref{eq:TN} is valid in the laboratory $xy$ basis, it must be transformed to the $sp-$mode basis, which is the proper eigenmode basis for isotropic media. The total transfer matrix is hence

\begin{equation}
\underline{\pmb{M}}=\underline{\pmb{S}}_0^{-1}\cdot\underline{\pmb{T}}_N\cdot\underline{\pmb{S}}_{0}
\label{eq:M}
\end{equation}
where $\underline{\pmb{T}}_N$ is given by \cref{eq:TN}. For the general case considered here it is not possible to give an explicit closed expression for the matrix $\underline{\pmb{M}}$. As a test case, let us consider the limit of an infinitely thick slab. This limit does not directly and algebraically follow from \cref{eq:M} (see supplemental material, section III for details). It can be shown that for a semi-infinite slab, the transfer matrix simplifies to

\begin{equation}
\underline{\pmb{M}}=\underline{\pmb{S}}_0^{-1}\cdot\underline{\pmb{S}}_1.
\label{eq:Minf}
\end{equation}

\subsection{Lifshitz formula}\label{sec:Fresnel}

{The Casimir-Lifshitz interaction free energy can be obtained from the secular determinant of the modes,  whose zeros give the eigenfrequencies of the bound states. \cite{Veble2009} However, in its turn the secular determinant itself can be rewritten in terms of the Fresnel reflection coefficients that can be obtained from the transfer matrix formalism. This route then connects the transfer matrix formalism with the Casimir-Lifshitz interactions which we elucidate in what follows.}
 {\sl Grosso modo} one can follow either the approach based on the Green function tensors, closest to the original papers by Lifshitz et al \cite{Lifshitz55,*Lifshitz61}. (See e.g. Refs. \cite{Schwinger78, Lambrecht2006, Scheel2008}) or follow the {\sl heuristic derivation} by van Kampen et al. \cite{vanKampen68, Schram1973}, which was later shown to be actually exact \cite{Barash_1984}, and based on the summation over the allowed EM modes  within a given geometry, more akin to the original paper by Casimir \cite{Casimir48}. It can be shown that both approaches lead to the same result, \cite{Intravaia2012} namely the Lifshitz formula for dispersion interactions. Here we will take advantage of this result.


We assume that the matrix $\underline{\pmb{M}}$ from \cref{eq:M} is known and that its entries are given by $\underline{\pmb{M}}= \Big( M_{ik}\Big)$ with $i,k\in\{1,2,3,4\}$.  In such a case, the Fresnel reflection matrix elements can be shown to be given by \cite{Yeh1979,Passler2017}:

\begin{subequations}\label{eq:Fresnel}
\begin{gather}
r_{ss}=\frac{M_{21} M_{33}-M_{23} M_{31}}{M_{11} M_{33}-M_{13} M_{31}}
\label{eq:rss}\\
r_{sp}=\frac{M_{33} M_{41}-M_{31} M_{43}}{M_{11} M_{33}-M_{13} M_{31}}
\label{eq:rsp}\\
r_{ps}=\frac{M_{11} M_{23}-M_{13} M_{21}}{M_{11} M_{33}-M_{13} M_{31}}
\label{eq:rps}\\
r_{pp}=\frac{M_{11} M_{43}-M_{13}M_{41}}{M_{11} M_{33}-M_{13} M_{31}}.
\label{eq:rpp}
\end{gather}
\end{subequations}
We have verified that combining \cref{eq:Minf}  with   \cref{eq:Fresnel} restores the known Fresnel reflection matrix \cite{Lekner1991, Broer2019} for a semi-infinite birefringent plate.

To determine the Casimir energy and torque, \cref{eq:rss,eq:rsp,eq:rps,eq:rpp} must be inserted into the Lifshitz formula. The energy per unit area is given by: 

\begin{equation}
 \frac{E_{Cas}}{A}=\frac{k_b T }{4\pi^2}\sum_{n=0}^{\infty}(1-\tfrac{1}{2}\delta_{0n})\int\limits_0^\infty\int\limits_0^{2\pi}\ln[D(k_\rho,\eta,\{\theta_j\}, \varphi, i\zeta_n)]k_\rho\ud k_\rho\ud\eta
\label{eq:Lifshitz}
 \end{equation}
where \cite{Lambrecht2006}
\begin{equation}
 D=\det(\underline{\pmb{I}}-\underline{\pmb{r}}_1(\{\theta_j\}, i\zeta_n)\cdot\underline{\pmb{r}}_2(\theta_1+\varphi, i\zeta_n)e^{-2k_3a})
\end{equation}
and the  reflection matrices are given by

\begin{equation}
 \underline{\pmb{r}}_q=\begin{pmatrix}
                        r_{q,ss}&r_{q,sp}\\
                        r_{q,ps}&r_{q,pp}
                       \end{pmatrix} \qquad q\in \{1,2\}.
\end{equation}
The subscript $q$ is merely a label for each multilayer stack. The entries of the matrices are given by \cref{eq:rss,eq:rsp,eq:rps,eq:rpp}. All quantities are evaluated at the imaginary Matsubara frequencies $ \zeta_n = \frac{2\pi n k_b T}{\hbar},$ so that each contribution to the Casimir energy decreases monotonically with $n$. The Casimir torque is then given by 

\begin{equation}\label{eq:torque}
 \tau(a,\varphi)=-\frac{\partial E_{Cas}}{\partial \varphi} 
\end{equation}
where $\varphi$ denotes the  angle between the optic axes of the layers of each stack closest to the gap. This result is equivalent to that of Ref. \cite{Veble2009} which uses the formalism of Berreman \cite{Berreman1972}, while here the same formalism has been used with the basis suggested by Yeh \cite{Yeh1979}.

\section{Modeling a Cholesteric liquid crystal}\label{sec:CLC}

\begin{figure}[!htbp]
\includegraphics[width=0.8\textwidth]{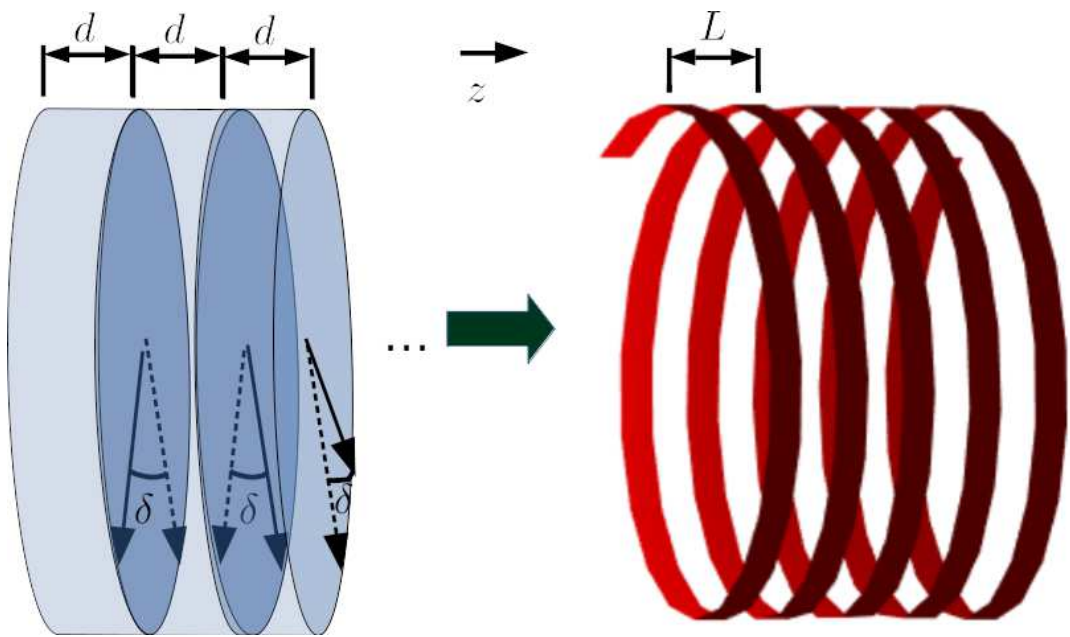}
 \caption{Illustration of the continuum approximation used here to model a liquid crystal. It consists of layers with identical thicknesses $d$ but each with a slightly different orientation, each with identical increments $\delta\ll1$. Consequently, the multilayer structure manifests as a continuous inhomogeneity in the direction perpendicular to the plane of reflection (defined as the $z$-direction), where the orientation effectively behaves as a helix with pitch length $L$. }\label{fig:NLC}
\end{figure}

A cholesteric liquid crystal is modeled as a nonmagnetic uniaxial planar multilayer stack.   All anisotropic layers of equal thickness $d$ are made of identical materials, and only differ by their orientation $\theta_j$. We start by writing  \cref{eq:ChangeOfBasis} as

\[\exp({-i\underline{\pmb{Q}}_j d})= \underline{\pmb{S}}_j\cdot\underline{\pmb{P}}_j\cdot\underline{\pmb{S}}_j^{-1}\]
for a single slab with label $j$.  

{Now let us formulate the conditions of validity of this approximation. In accordance with an actual cholesteric liquid crystal \cite{Ryabchun2018}, each layer is assumed to be infinitesimally thin, i.e. $d\rightarrow0$. Moreover, let the  difference in orientation between each subsequent layer, $\delta$, also be infinitesimally small, in such a way that this quantity parametrically describes a helix in space. Consequently, an infinitesimal change in $z$ within the liquid crystal is proportional to an infinitesimal change in the orientation: } 

\begin{equation}\label{eq:helix}
    \ud z=d=\frac{L}{2\pi} \delta,
\end{equation}
{where $L$ denotes the pitch length, in which the orientations of the layers complete one period, and $\delta\ll1$. Hence there is some freedom of choice, we can use either $z$ or $\theta$ as a (continuous) variable. In what follows we will use two variables interchangeably. Another assumption is that the total thickness $Nd$, where $N$ denotes the total number of layers, is infinite. This is justifiable because a typical thickness of  liquid crystals is of the order of several microns, and we will limit ourselves to separation distances of less than one micron ($a\ll Nd$), where the Casimir-Lifshitz torque is likely to be large enough to be measured \cite{Somers2018}. Finally, it will be assumed that the pitch length  of the cholesteric liquid crystal $L$, in which the orientations of the layers complete one period, is smaller than the total thickness. Hence at least one period is assumed to fit inside the crystal. We can summarize the condition for our approximation as follows:}

\begin{equation}\label{eq:condition}
    d\ll L \ll Nd,
\end{equation}
which means that the difference between the first orientation and last orientation is at least $2\pi$ radians. This is applicable to a cholesteric liquid crystal \cite{Ryabchun2018}. We also require $d$ to be small on the scale of the matrix norm of $\underline{\pmb{Q}}_j$:

\begin{equation}
||\underline{\pmb{Q}}_j||d\ll1\quad\text{for all } 1\leq j\leq N.
\label{eq:BCHCond}
\end{equation}
Physically, \cref{eq:BCHCond} can be understood as the thickness $d$  being much smaller than the typical values of the relevant wavelengths.   After all, in a continuous medium approach, the relevant wavelengths are assumed to be larger than the typical distances between atoms. Since the physical thickness is actually of the same order of magnitude, it can be approximated as a numerical integration stepsize in the context of a continuous medium approach.  The condition \cref{eq:BCHCond} allows us to use the Baker-Campbell-Hausdorff (BCH) formula \cite{HighamBook2008}. The second condition of \cref{eq:condition}  allows us to use \cref{eq:Minf} instead of \cref{eq:M}.

In section IV of the supplemental material we derive the following expression valid up to the second order in the angular deviation in the orientation between the neighboring uniaxial layers
\begin{equation}
\prod_{j=1}^{N}\exp\left(-id\underline{\pmb{Q}}_{j}\right)=\exp\Bigg(-id\sum_{j=1}^{N}\underline{\pmb{Q}}_{j}+O(\delta^2)\Bigg)\longrightarrow\exp\Bigg(-i\int\limits_{\min(z)}^{\max(z)}\underline{\pmb{Q}}(\theta)\ud z + O(\delta^2)\Bigg),
\label{eq:NLayerDiscrete}
\end{equation}
where eventually we assumed that $j$ changes as a continuous variable associated with $z$. The full derivation is somewhat lengthy so we will summarize it here. In short, we use the Baker-Campbell-Haussdorff (BCH) formula \cite{HighamBook2008} up to third order in $\delta$. Then we combine the BCH formula with finite difference coefficients \cite{Fornberg1988} to determine that, three nearest neighbors to the right are needed to obtain the same accuracy as the BCH formula.

In order to evaluate the argument of the exponential matrix in \cref{eq:NLayerDiscrete}, and we will try to determine its diagonalized form. Let us focus on the leading order term: $-id\sum_{j=1}^{N} \underline{\pmb{Q}}_{j}$. Even though \cref{eq:ChangeOfBasis} tells us that each individual matrix $\underline{\pmb{Q}}_{j}$ is diagonalizable, it does not follow in general that their sum is also diagonalizable. However, within this particular approximation it is assumed that the commutators between the matrices $\underline{\pmb{Q}}_{j}$ are negligible and the matrices effectively commute, with the error of this approximation being of second order. Commuting  diagonalizable matrices can be shown to have the same eigenvectors. Consequently, each diagonalizable matrix $\underline{\pmb{Q}}_{j}$ can be transformed to the same eigenvector basis via one matrix, say $\underline{\pmb{S}}$. In other words, the matrices $\underline{\pmb{Q}}_{j}$ can be simultaneously diagonalized:
\[\sum_{j=1}^{N}\underline{\pmb{Q}}_{j} = \underline{\pmb{S}}\cdot\underline{\pmb{D}}\cdot\underline{\pmb{S}}^{-1}\]
where $\underline{\pmb{D}}$ is a diagonal matrix. This furthermore implies that:

\begin{equation}
\exp\left(-id\sum_{j=1}^{N}\underline{\pmb{Q}}_{j}\right)=\underline{\pmb{S}}\exp\left(-id\underline{\pmb{D}}+O(d^2)\right)\underline{\pmb{S}}^{-1}.
\label{eq:Simultaneous}
\end{equation}
 We know that the matrices $\underline{\pmb{S}}$ and $\underline{\pmb{D}}$ exist but now we want to determine their explicit form.  The physical interpretation of \cref{eq:Simultaneous} is that, in this approximation, the orientations of the different layers are averaged over the entire crystal. Since $d\rightarrow0$ and $N\rightarrow\infty$, it can be proposed that $d\sim1/N$, and the left hand side of \cref{eq:Simultaneous} becomes an average over all eigenvalues. Hence it follows that

\begin{equation}
\underline{\pmb{D}}=\frac{1}{N}\sum_{j=1}^{N}\diag(q_{e}((\theta_j-\eta)),-q_{e}((\theta_j-\eta)),q_o,-q_o)
\label{eq:DDiscAvg}
\end{equation}
which changes to an average with respect to a continuous distribution where the sum is replaced by an integral. The approximation of \cref{eq:DDiscAvg} is of course simple and crude, its error being of second order in the stepsize $d$. It assumes that the planar multilayered system can be approximated as a single slab with an averaged orientation. Essentially the effect of the presence of multiple layers is treated additively in this case. 

The transfer matrix for such a configuration is given by  (see \cref{eq:Simultaneous}):

\begin{equation}
\exp\left(-id\sum_{j=1}^{N}\underline{\pmb{Q}}_{j}\right)=\left<\underline{\pmb{S}}\right>\underline{\pmb{P}}_{tot}(\left<q_e\right>)\left<\underline{\pmb{S}}\right>^{-1},
\label{eq:avgSlab}
\end{equation} 
where
\begin{equation}
\underline{\pmb{P}}_{tot}(\left<q_e\right>)=\diag(\exp(-id_{tot}\left<q_e\right>),\exp(id_{tot}\left<q_e\right>,\exp(-id_{tot}q_o), \exp(id_{tot}q_o)).
\label{eq:Ptot}
\end{equation}
Here $d_{tot}=Nd$ represents the total thickness, and in the limit $d_{tot} \rightarrow \infty$,  $\underline{\pmb{P}}_{tot}$ tends to {${\underline{\pmb{P}}_{tot}\rightarrow\diag(1,0,1,0)}.$} {(This does not follow directly algebraically, but from physical considerations, see Supplemental Material, section III ).} In \cref{eq:Ptot} the eigenvectors are the ones corresponding to $\left<q_e\right>$:

\begin{equation}
\left<\underline{\pmb{S}}\right>=\underline{\pmb{S}}(\left<q_e\right>,\left<\theta\right>),
\label{eq:Savg}
\end{equation}
where $\underline{\pmb{S}}$ is given by \cref{eq:SDef} and  $\left<q_e\right>$ denotes the extraordinary eigenvalue, averaged over the orientations:
\begin{equation}
 \left<q_e\right>=\frac{1}{N}\sum_{j=1}^{N}q_{e}((\theta_j-\eta)) \longrightarrow \frac{1}{N\delta}\int\limits_0^{N\delta}q_{e}(\theta-\eta)\ud\theta.
\label{eq:qeavg}
\end{equation}
The rightmost expression follows after the continuity approximation that allows us to replace the sums over the discrete index $j$ by integrals over the continuous variable $\theta$.  We assume here that the pitch length, the length of one period, is {smaller than the total thickness, see \cref{eq:condition}}.

{Due to the periodicity of $q_e$,  $N\delta$ can be replaced by $\pi$, so that $q_{e}$ is averaged over one period and it can take}\\ {all possible real values}. {This is because the thickness of the liquid crystal is assumed to be larger than the pitch length.  (See \cref{eq:condition}) Consequently, at least one period can be fit into the crystal. The integer part of the ratio between the thickness and the pitch length will dominate the remainder, because the remainder represents a part of the liquid crystal located at an infinite distance from the plane of reflection. Setting $N\delta=\pi$}  also allows us to set $\eta=0$, because the integral becomes translation invariant over one period. In this case, the integral of \cref{eq:qeavg} can be evaluated analytically:

\begin{equation}
  \left<q_e\right>=\frac{1}{\pi}\int\limits_0^{\pi}q_{e}(\theta)\ud\theta=\pm\frac{2}{\pi}\sqrt{\varepsilon_{1x}\left(\tfrac{k_\rho^2}{\varepsilon_{1y}}+\tfrac{\zeta^2}{c^2}\right)}E\left(\frac{k_\rho^2(\tfrac{\varepsilon_{1x}}{\varepsilon_{1y}}-1)}{\varepsilon_{1x}\left(\tfrac{k_\rho^2}{\varepsilon_{1y}}+\tfrac{\zeta^2}{c^2}\right)}\right),\label{eq:qeavg3}
\end{equation}
where $E(z)$ denotes the complete elliptic integral of the second kind, given by \cite{AbramowitzStegun}

\[E(z)=\int_{0}^{\tfrac{\pi}{2}}\sqrt{1-z\sin^2t}\ud t,\]
the value of which is real for arguments $-1\leq z\leq1$.

To determine the average orientation, we must solve the following equation:
\begin{equation}\label{eq:qeavgEq}
  \left<q_e\right>=\pm\sqrt{\varepsilon_{1x}\tfrac{\zeta^2}{c^2}+\tfrac{\varepsilon_{1x}}{\varepsilon_{1y}}k_\rho^2\cos^2\left<\theta\right>+k_\rho^2\sin^2\left<\theta\right>}  
\end{equation}
which  has four possible solutions:

\begin{equation}\label{eq:thsols}
 \left<\theta\right>=\pm\arccos\left(\pm\frac{\sqrt{k_\rho^2+\varepsilon_{1x}\zeta^2/c^2-\left<q_e\right>^2}}{k_\rho\sqrt{1-\tfrac{\varepsilon_{1x}}{\varepsilon_{1y}}}}\right).   
\end{equation}
Here the different signs of $\left<\theta\right>$ represent the different chiralities of the liquid crystal: a positive $\left<\theta\right>$ is associated with right handed chirality and the minus sign corresponds to left handed chirality, if the axial wave vector component is positive. The different signs in the argument of the arc-cosine in \cref{eq:thsols} come from the different square roots in \cref{eq:qeavgEq},  associated with the two different propagation directions of the extraordinary modes in the liquid crystal. {However, since the quantity $\left<q_e\right>$ is an average along the helix, its sign must be interpreted as the direction of the helical spiral. Alternatively, the different signs of the square root of  $\left<q_e\right>$ in \cref{eq:qeavgEq} can be thought of as the location of the crystal: since the thicknesses of the crystals are assumed to be infinite, the negative square root represents the crystal being on the left and the positive square root represent the crystal being on the right. This distinction depends on the arbitrary choice of coordinates, and it is hence not physical.  } \cref{eq:thsols} provides four solutions whereas there are only two physically distinguishable shapes of the helix, namely left- and right handed. Therefore we need to establish how they can be physically interpreted. {We require that the physical $\left<\theta\right>$ is real. } First we have to verify that such solutions exist. {To this end we distinguish two cases: $\varepsilon_{1y}\geq\varepsilon_{1x}$ and $\varepsilon_{1y}\leq\varepsilon_{1x}$.}  In order for $\left<\theta\right>$ to be real, we need to check that 

\[0\leq k_\rho^2+\varepsilon_{1x}\tfrac{\zeta^2}{c^2}-\left<q_e\right>^2\leq k_\rho^2\left(1-\tfrac{\varepsilon_{1x}}{\varepsilon_{1y}}\right),\]
which can be seen to be the case since 

\[\min\left(1,\tfrac{\varepsilon_{1x}}{\varepsilon_{1y}}\right)\leq\tfrac{\varepsilon_{1x}}{\varepsilon_{1y}}\cos^2\theta+\sin^2\theta\leq\max\left(1,\tfrac{\varepsilon_{1x}}{\varepsilon_{1y}}\right)\]
for all real $\theta$. {Here it is assumed that $|\left<\theta\right>|\leq\pi$. } Hence the real value of the average within the domain $-\pi\leq\left<\theta\right>\leq\pi$ orientation is
\begin{subequations} \label{eq:thavg_left} 
    \begin{equation}
        \left<\theta\right>_{L}=  
                       \arccos\left(-\frac{\sqrt{k_\rho^2+\varepsilon_{1x}\zeta^2/c^2-\left<q_e\right>^2}}{k_\rho\sqrt{1-\tfrac{\varepsilon_{1x}}{\varepsilon_{1y}}}}\right)\\\label{eq:thetaLL}
    \end{equation}

    \begin{equation}
        \left<\theta\right>_{R}=                       -\arccos\left(-\frac{\sqrt{k_\rho^2+\varepsilon_{1x}\zeta^2/c^2-\left<q_e\right>^2}}{k_\rho\sqrt{1-\tfrac{\varepsilon_{1x}}{\varepsilon_{1y}}}}\right)\\\label{eq:thetaLR}
    \end{equation} 
\end{subequations}
if the crystal is on the left, and
\begin{subequations} \label{eq:thavg_right} 
    \begin{equation}
        \left<\theta\right>_{L}=  
                       -\arccos\left(\frac{\sqrt{k_\rho^2+\varepsilon_{1x}\zeta^2/c^2-\left<q_e\right>^2}}{k_\rho\sqrt{1-\tfrac{\varepsilon_{1x}}{\varepsilon_{1y}}}}\right)\\\label{eq:thetaRL}
    \end{equation}
    \begin{equation}
        \left<\theta\right>_{R}=                       \arccos\left(\frac{\sqrt{k_\rho^2+\varepsilon_{1x}\zeta^2/c^2-\left<q_e\right>^2}}{k_\rho\sqrt{1-\tfrac{\varepsilon_{1x}}{\varepsilon_{1y}}}}\right)\\\label{eq:thetaRR}
    \end{equation}  
\end{subequations}
if the crystal is on the right. Here  $\left<q_e\right>$ is given by \cref{eq:qeavg3}. The subscripts $L$ and $R$ in \cref{eq:thavg_right,eq:thavg_left}  denote the left- and  right handedness of the liquid crystal, respectively. Note that \cref{eq:thavg_right,eq:thavg_left} depend on frequency, which in turn affects the frequency dependence of the reflection matrix. Finally, \cref{eq:qeavg3,eq:thavg_right,eq:thavg_left} have to be inserted into \cref{eq:SDef} in order to obtain $\left<\underline{\pmb{S}}\right>_1$.

Now we move on to the next order term in the BCH expansion. Here the contribution of nested commutators is neglected, i.e. commutators are assumed to commute with other matrices. In order to take advantage of \cref{eq:avgSlab} we apply the BCH formula again to separate the BCH terms as follows (see Supplemental material, section IV for a detailed derivation):

 \begin{equation}
\exp\left(-i\frac{L}{\pi}\int\limits_0^{N\delta}\underline{\pmb{Q}}(\theta)\ud\theta-3\frac{L^2}{\pi^2}\int\limits_{0}^{(N-3)\delta}\left[\underline{\pmb{Q}}(\theta),\frac{\partial \underline{\pmb{Q}}(\theta)}{\partial \theta}\right]\ud\theta+O(\delta^3)\right)
 \label{eq:separated}
 \end{equation}
\[=\left<\underline{\pmb{S}}\right>_1\underline{\pmb{P}}_{tot}(\left<q_e\right>)\left<\underline{\pmb{S}}\right>_1^{-1}\exp\Bigg(-3\frac{L^2}{\pi^2}\int\limits_{j=1}^{(N-3)\theta}\left[\underline{\pmb{Q}}(\theta),\frac{\partial \underline{\pmb{Q}}(\theta)}{\partial \theta}\right]\ud\theta\Bigg),\]
where we have taken advantage of the transfer matrix for the single interface case, \cref{eq:Minf}.
The argument of rightmost exponential matrix on the right hand side of \cref{eq:separated} is not jointly diagonalizable with the first BCH term. \footnote{The matrices may still be approximately jointly diagonalizable. However at this point it is not clear how to estimate this approximation, or what the common eigenvector basis should be.} However, this correction is expected to be small compared to the averaged factor on its left. This allows us to write it as a Taylor expansion. Hence we obtain

\begin{equation}
\
 \prod_{j=1}^{N}\exp\left(-id\underline{\pmb{Q}}_{j}\right)=\left<\underline{\pmb{S}}\right>_1\underline{\pmb{P}}_{tot}(\left<q_e\right>)\left<\underline{\pmb{S}}\right>_1^{-1}\left(\underline{\pmb{I}}-3\frac{L^2}{\pi^2}\int\limits_0^{(N-3)\delta}\left[\underline{\pmb{Q}}(\theta),\frac{\partial \underline{\pmb{Q}}(\theta)}{\partial \theta}\right]\ud\theta+O(\delta^3)\right).
\label{eq:Taylor}
\end{equation}
The leading order term in \cref{eq:Taylor} can be understood as a slab with infinite thickness and an averaged orientation. The second order correction describes the effect due to the finite thicknesses of the layers and their slightly different orientations. For a liquid crystal of infinite thickness \cref{eq:Taylor} becomes

\begin{equation}
 \prod_{j=1}^{N}\exp\left(-id\underline{\pmb{Q}}_{j}\right)=\underline{\pmb{S}}_0^{-1}\left<\underline{\pmb{S}}\right>_1\left(\underline{\pmb{I}}-3\frac{L^2}{\pi^2}\int\limits_0^{(N-3)\delta}\left[\underline{\pmb{Q}}(\theta),\frac{\partial \underline{\pmb{Q}}(\theta)}{\partial \theta}\right]\ud\theta+O(\delta^3)\right),
\label{eq:TaylorInfThick}
\end{equation}

Next we are faced with the task of writing \cref{eq:TaylorInfThick} as an expansion in $\delta$. Note that the integral absorbs one factor of $\delta$ and that the integration from 0 to $N\delta=\pi$ will not contribute to the integral (See supplemental material, section I for the explicit expressions for the commutator), so the integral is expected to be small $\sim\delta$. Within the approximations used here, it suffices to calculate the integral up to first order in $\delta$. (A more precise calculation would be beyond the approximation and hence be inconsistent). We will denote this first order coefficient by $\underline{\pmb{A}}$. Hence the transfer matrix is

\begin{equation}
 \prod_{j=1}^{N}\exp\left(-id\underline{\pmb{Q}}_{j}\right)= \underline{\pmb{S}}_0^{-1}\left<\underline{\pmb{S}}\right>_1\left(\underline{\pmb{I}}-3\frac{L^2}{\pi^2}\underline{\pmb{A}}\delta+O(\delta^3)\right),
 \label{eq:final}
\end{equation}
where $\underline{\pmb{A}}$ is given by the integral in \cref{eq:TaylorInfThick}, and its explicit expressions can be found in the supplemental material, section I. Note that the error of the BCH formula is of third order in $\delta$, as the higher order term \cref{eq:final} is actually quadratic in $\delta$. So this term does not have to follow the sign convention of \cref{eq:thavg_left,eq:thavg_right}. Since the approximate transfer matrix is known now, \cref{eq:Fresnel} can be applied to it to obtain the corresponding Fresnel matrix. Note that the factor of $L^2$ cancels out of the higher order correction for the Fresnel matrix. However, it's worth pointing out that $\delta$ is inversely proportional to $L$, (see \cref{eq:helix}) so that the final result still depends on the pitch length. We again refer to the supplemental material, section I for the explicit expressions for the corrections to the Fresnel matrix. It's worth noting that the Fresnel matrices are invariant under $\pmb{k}\rightarrow-\pmb{k}$, hence they are identical whether they are placed on the left or on the right (if they have the same orientation). However in general this is not necessarily true \cite{Bing2021}. 

\section{Results}\label{sec:Results}
\begin{figure}[!htbp]
\includegraphics[width=0.49\textwidth]{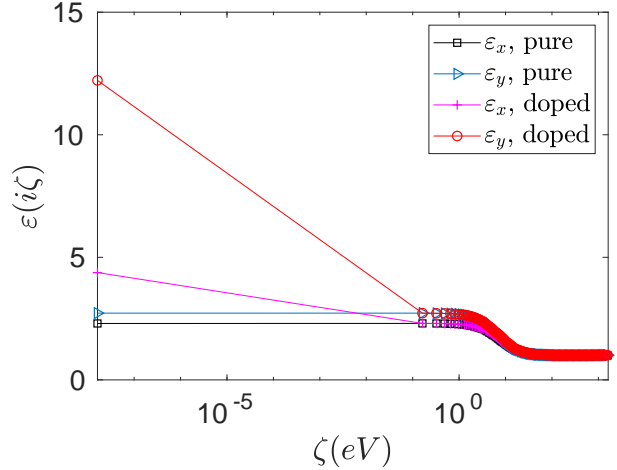}
 \caption{Dielectric functions of pure, nematic and doped, cholesteric 5CB. The dielectric function of nematic 5CB as proposed in Ref. \cite{Kornilovitch2012, Kornilovitch2018} (indicated by the black squares and blue triangles) is modified to account for the presence of a chiral dopant. This affects the static values (see the magenta crosses and the red circles), which were obtained from Ref. \cite{Kocakuelah2021}  }\label{fig:5CBeps}
\end{figure}
{Here we will use the results derived  in section \ref{sec:CLC} to calculate the Casimir-Lifshitz torque given by \cref{eq:torque}. The first configuration we consider is composed of a birefringent half-space, a finite vacuum gap of thickness 10 nm $<a<$1 $\mu$m, and a cholesteric half-space. Next we will investigate the combination of two cholesteric liquid crystals. The former in general should be accessible to experiments \cite{Somers2018}. While realistically, surface roughness generally plays a significant role at this range \cite{PeterRoughness}, the roughness effects have been considered elsewhere in detail \cite{d0paper, BroerEPL2011, BroerPRB2012} and we will ignore them here. We have verified our calculations by reproducing earlier results obtained by implementations of the Barash formula \cite{Munday2005, Somers2015, Thiyam2018} (See supplemental material, section V).}

{In order to calculate Casimir-Lifshitz interactions, the frequency dependent dielectric function of the interacting materials must be known. This is why we introduce it here first. We will ignore the temperature dependence of the dielectric function and assume a constant temperature of 293 K. This is because we are interested in the cholesteric phase only.}

{The dielectric function of a nematic liquid crystal can be described by a semi-empirical three band model \cite{Wu1991}:}

\begin{equation}\label{eq:eps5CB}
    \varepsilon_{N,x,y}(i\zeta)=\left(1+\sum_{k=1}^{3} \frac{C_{k,x,y}}{1+\tfrac{\zeta^2}{\omega_k^2}}\right)^2.
\end{equation}
{However,a pure 5CB crystal is nematic and not  cholesteric. It needs to be doped to reach a cholesteric phase, which is typically described  by a  Debye model \cite{delaFuente2014}:}

\begin{equation}\label{eq:epsD}
    \varepsilon_{D,x,y}(i\zeta)=\varepsilon_{\infty,x,y}+\frac{\varepsilon_{x,y}(0)-\varepsilon_{\infty,x,y}}{1+\zeta\tau}
\end{equation}
{Therefore we propose to model a doped nematic liquid crystal as the sum of these dielectric functions}

\begin{equation}\label{eq:Debye}
    \varepsilon_{x,y}(i\zeta)=\varepsilon_{D,x,y}(i\zeta)+\varepsilon_{N,x,y}(i\zeta)
\end{equation}
{However, the Debye relaxation processes take place on a much smaller frequency scale than the Matsubara frequencies at room temperature. More precisely, the smallest Matsubara frequency ($n=1$) at 293 K is six orders of magnitude larger than the typical values of the Debye relaxation frequencies. \cite{delaFuente2014,Kocakuelah2021} 
 Therefore we propose to approximate it as the static term from \cref{eq:Debye}:} 

\begin{equation}\label{eq:eps_tot}
\varepsilon_{x,y}(i\zeta)\approx\varepsilon_{D,x,y}(0)+\varepsilon_{5CB,x,y}(i\zeta)
\end{equation}
{in other words, the Debye processes only affect the static permittivity of the liquid crystal.} The different components of the dielectric functions for both the nematic and cholesteric case in  \cref{fig:5CBeps}.

\begin{figure}[!htbp]
\begin{subfigure}[t]{0.45\textwidth}
\centering
\includegraphics[width=\linewidth]{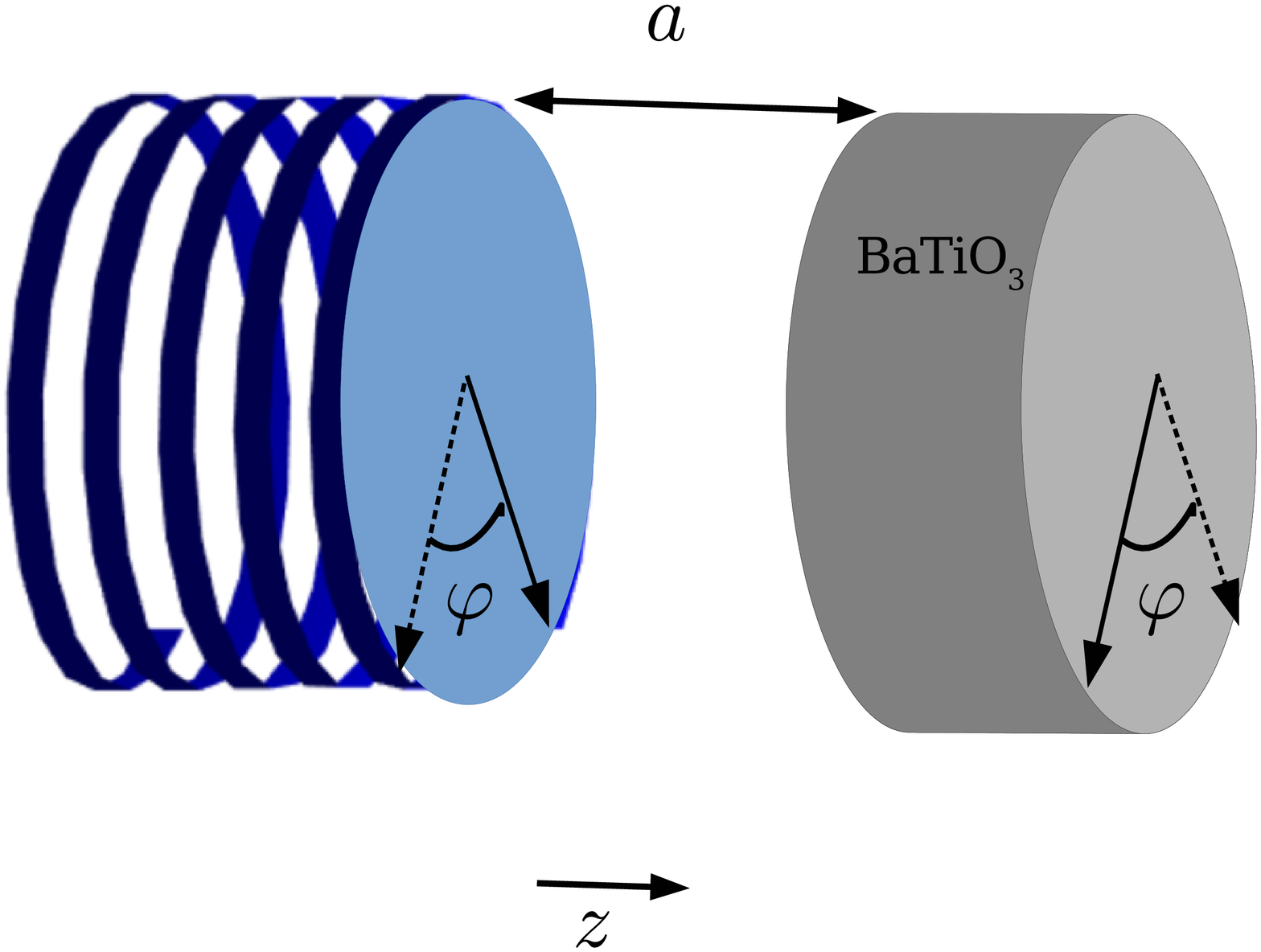}
 \caption{ }
 \label{fig:Left5CBBaTiO3}
\end{subfigure}
\hfill
\begin{subfigure}[t]{0.45\textwidth}
\includegraphics[width=\linewidth]{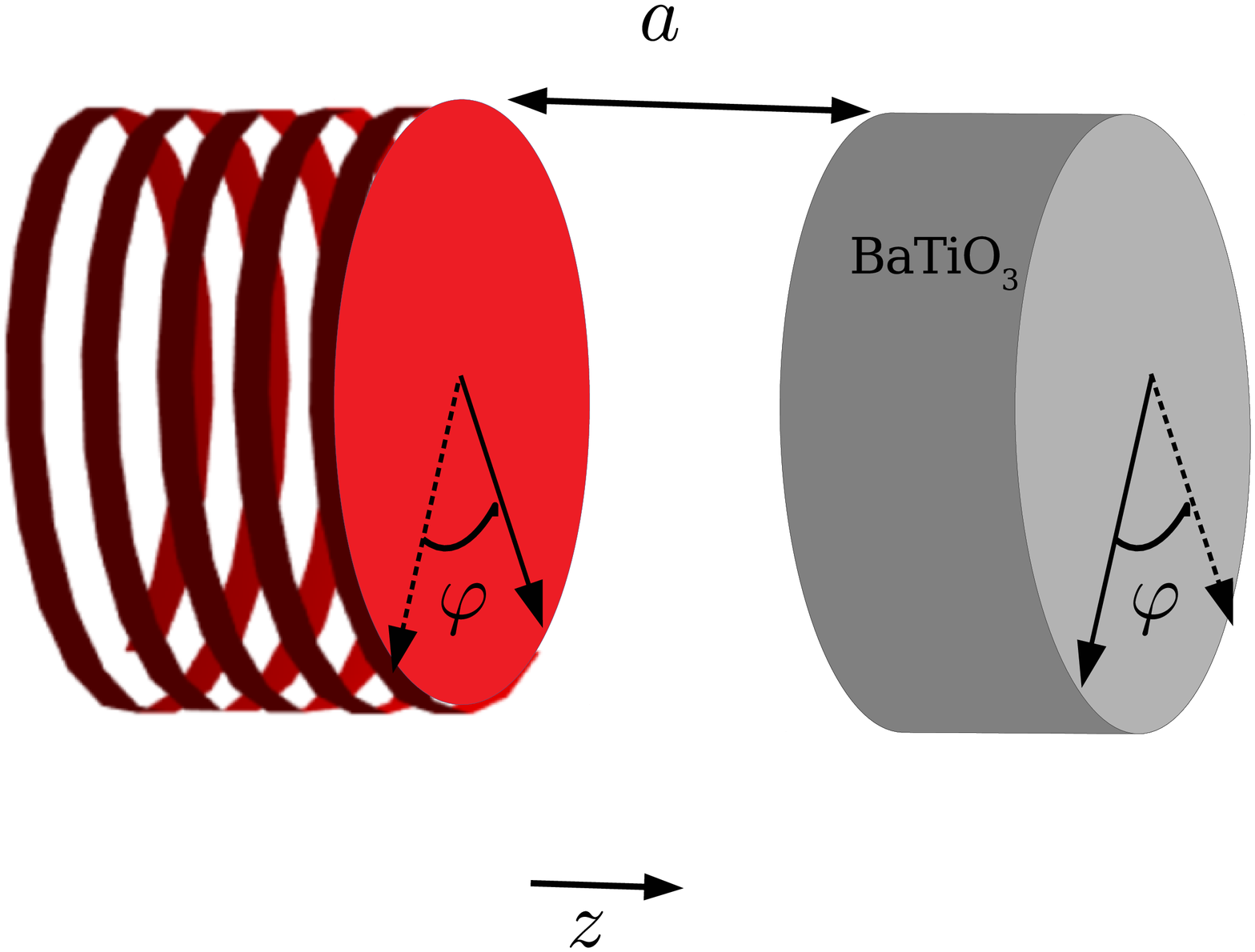}
 \caption{}
 \label{fig:Right5CBBaTiO3}
\end{subfigure}
\hfill
\begin{subfigure}[t]{0.45\textwidth}
\includegraphics[width=\linewidth]{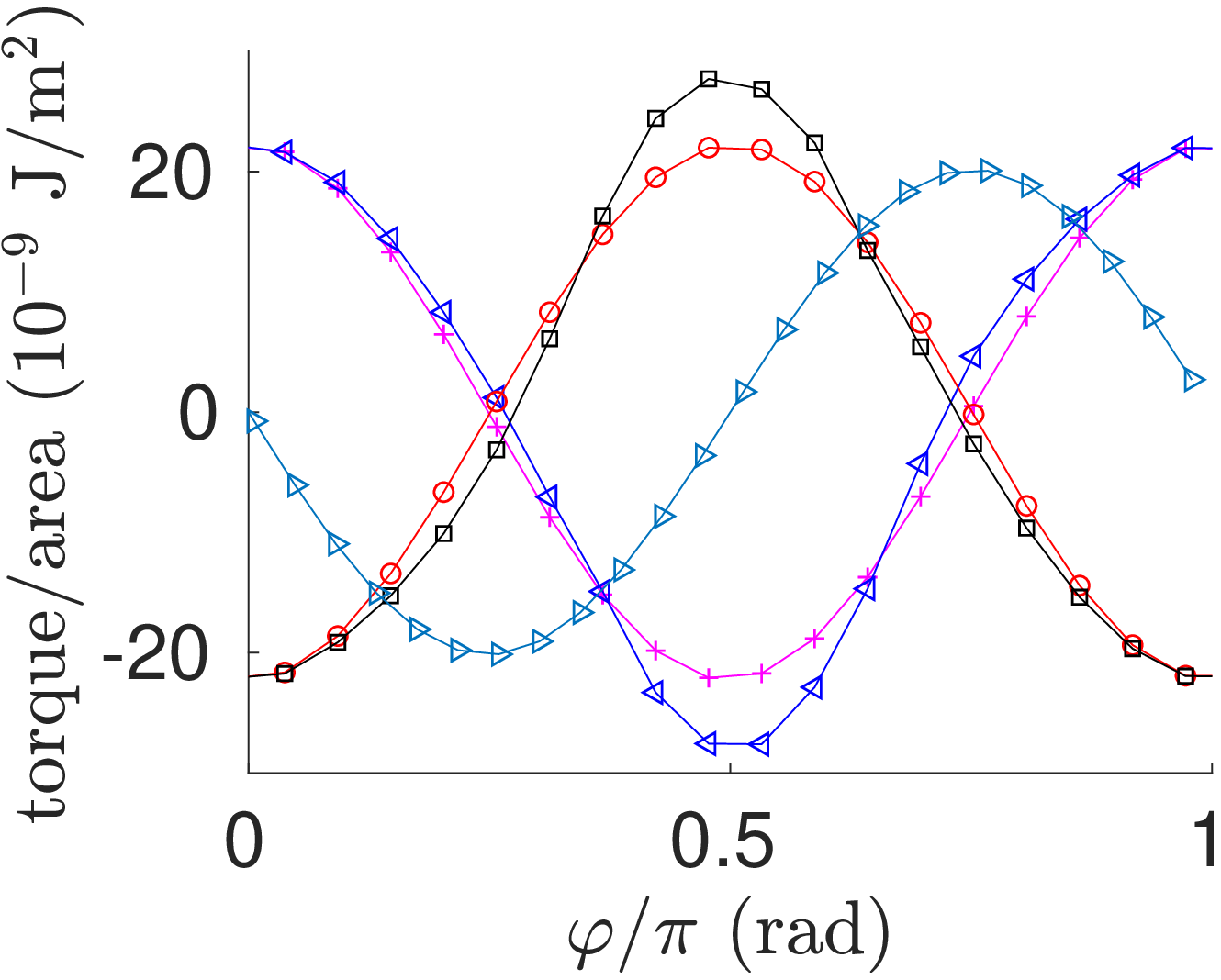}
 \caption{ }
 \label{fig:BaTiO35CBphi}
 \end{subfigure}
 \hfill
\begin{subfigure}[t]{0.45\textwidth}
\includegraphics[width=\linewidth]{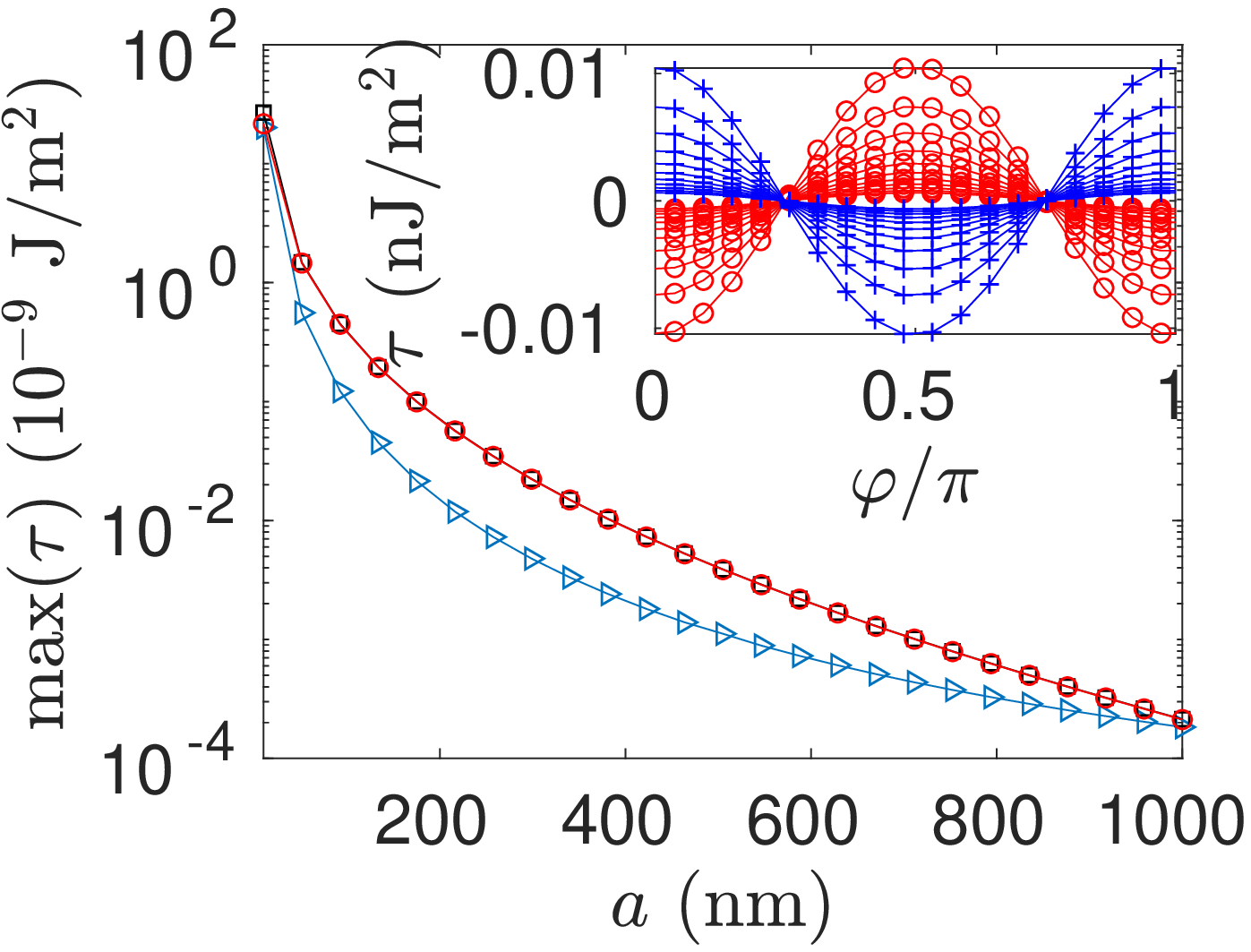}
 \caption{ }
 \label{fig:BaTiO35CBdistance}
 \end{subfigure}
 \caption{The Casimir torque between a birefringent (BaTiO$_3$) crystal and a cholesteric liquid crystal. (a): a left handed facing a birefringent one. (b): a right handed  crystal facing a birefringent one. (c): the Casimir torque as a function of misalignment angle at a distance of $a=10$ nm. The different curves are: the triangles pointing right show the usual result for a nematic liquid crystal. Red circles: left handed crystal, $\delta=0$. Blue triangles pointing left: left handed crystal, $\delta\neq0$. Magenta crosses: right handed crystal, $\delta=0$. Black squares: right handed crystal, $\delta\neq0$. (d): The amplitude of the Casimir torque as a function of separation. The triangles pointing right show the nematic case. The cholesteric case is shown by the other two curves, of which the red circles represents $\delta=0$ and the black squares show the perturbed case. The inset of panel (d) shows the Casimir torque for the cases (a) (red circles) and (b) (blue crosses) at distances between 380 and 800 nm. }\label{fig:BaTiO35CB}
\end{figure}

{Now we are in a position to specify the materials, i.e. assign values to the parameters in \cref{eq:eps_tot}.  Let the cholesteric liquid crystal consist of 96 mass \% nematic 5CB (4- cyano-4'-pentyl-biphenyl) doped with 4 mass \% chiral dopant S811. \cite{Kocakuelah2021} The value of the static Debye term for this mixture was taken from a recent experiment \cite{Kocakuelah2021}. The dielectric function of 5CB  was established in Ref. \cite{Kornilovitch2012} based on data from Ref. \cite{Wu1993}. The optical data of barium titanate (BTO $\rm  Ba Ti O_3$) \cite{BERGSTROM1997} will be used to describe the semi-infinite birefringent half-space. For more details about the parameter values and their effect on the Casimir-Lifshitz interaction we refer to section VI of the supplemental material.}  Throughout this section we choose the value of $\delta$ (if nonzero) to be 0.03, which corresponds to a pitch length of 200 nm. (See \cref{eq:helix}).

\begin{figure}[ht]
\begin{subfigure}[t]{0.45\textwidth}
\centering
\includegraphics[width=\linewidth]{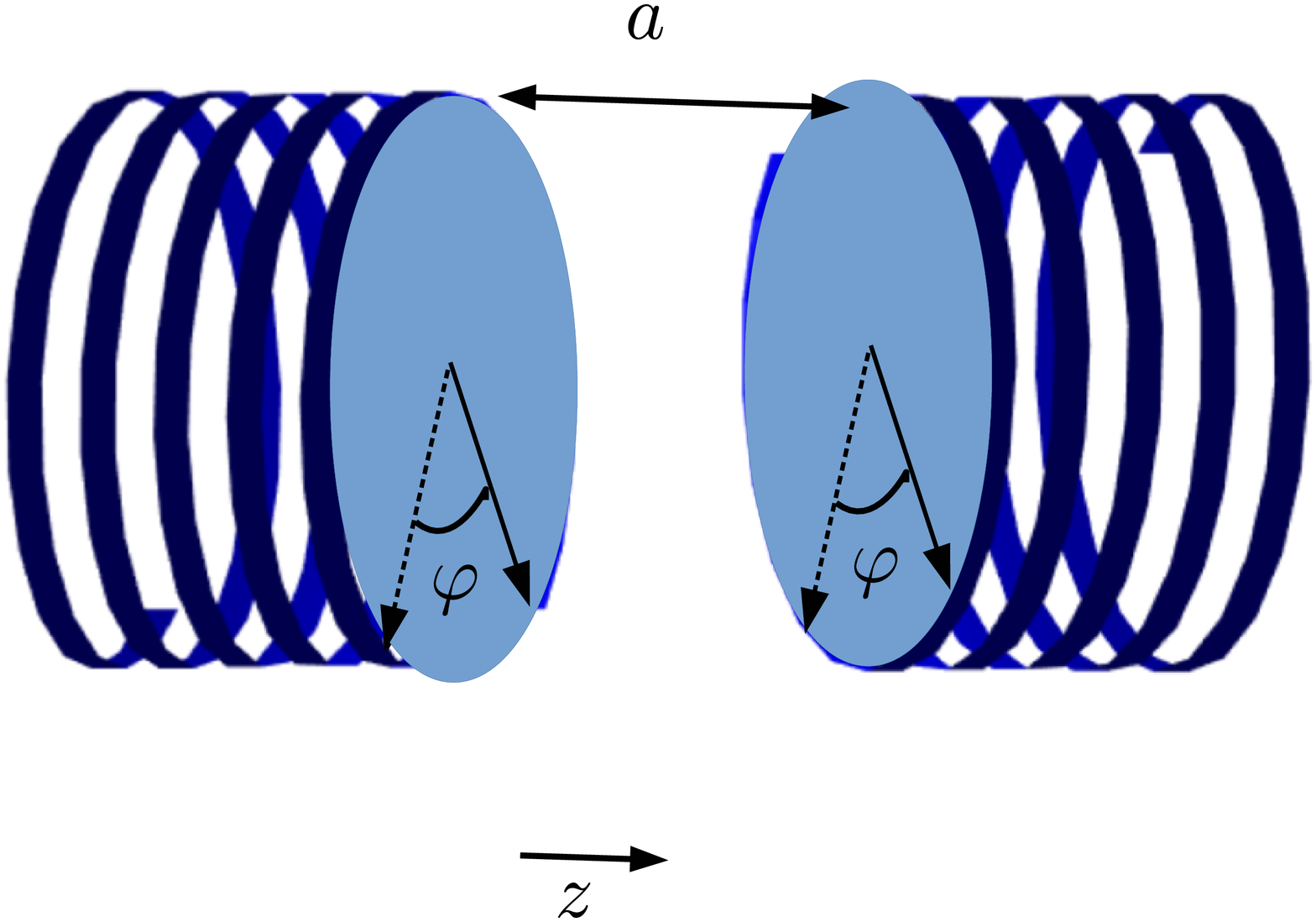}
 \caption{ }
 \label{fig:HomoChiralRight}
\end{subfigure}
\hfill
\begin{subfigure}[t]{0.45\textwidth}
\includegraphics[width=\linewidth]{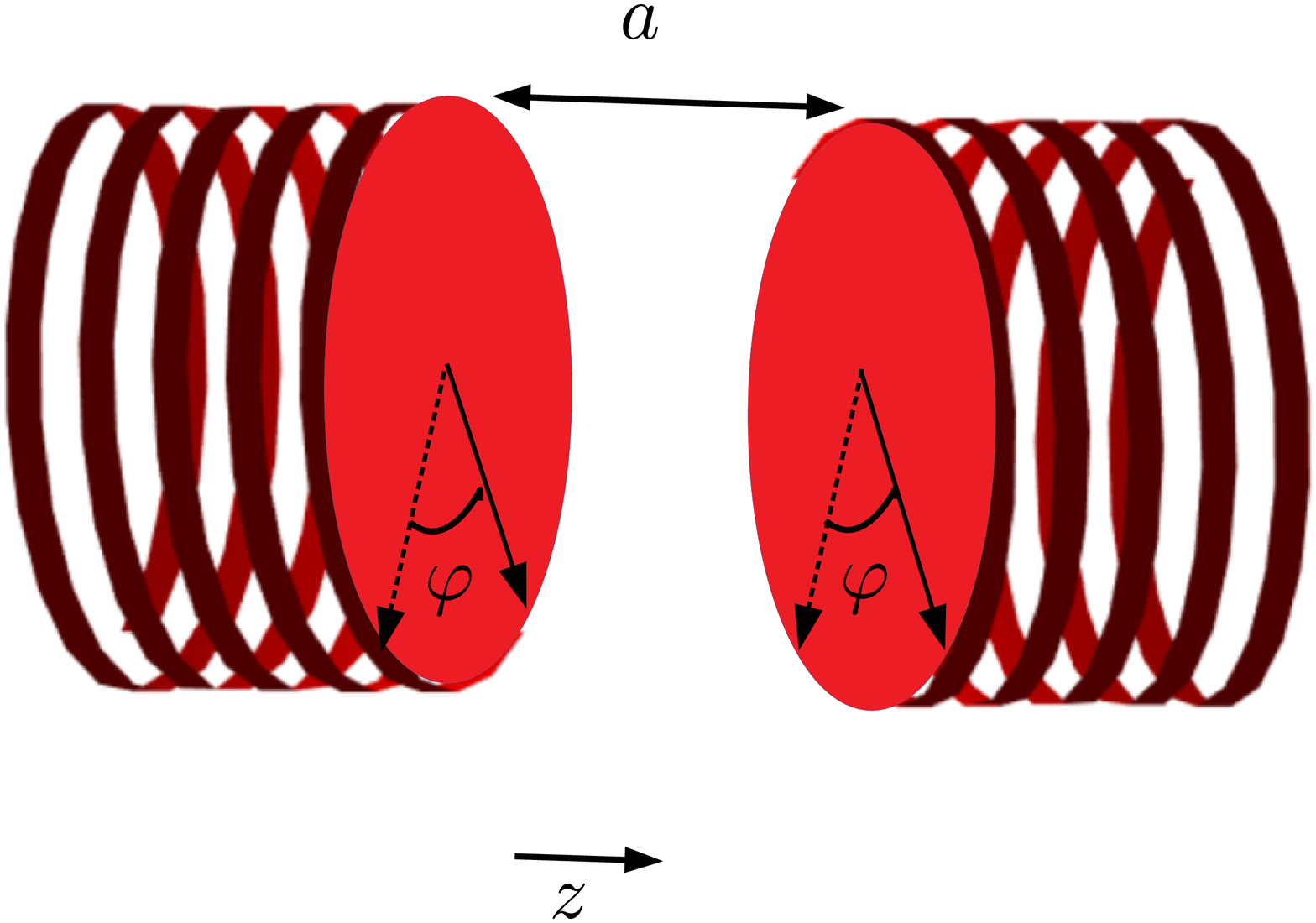}
 \caption{ }
 \label{fig:HomoChiralLeft}
\end{subfigure}
\hfill
 \begin{subfigure}[t]{0.45\textwidth}
\includegraphics[width=\linewidth]{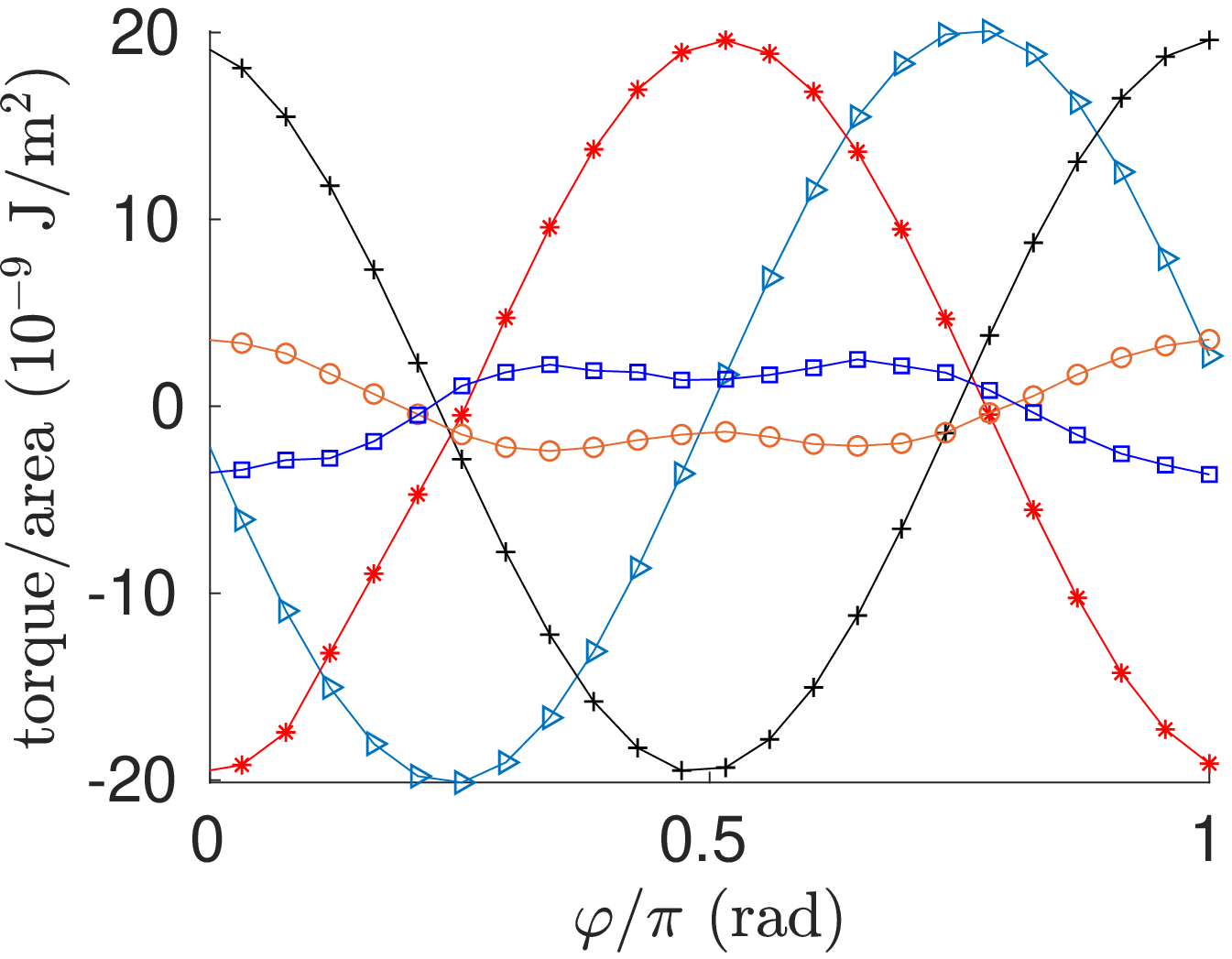}
\caption{ }
\hfill
\label{fig:HomoChiralPhi}
 \end{subfigure}
  \begin{subfigure}[t]{0.45\textwidth}
\includegraphics[width=\linewidth]{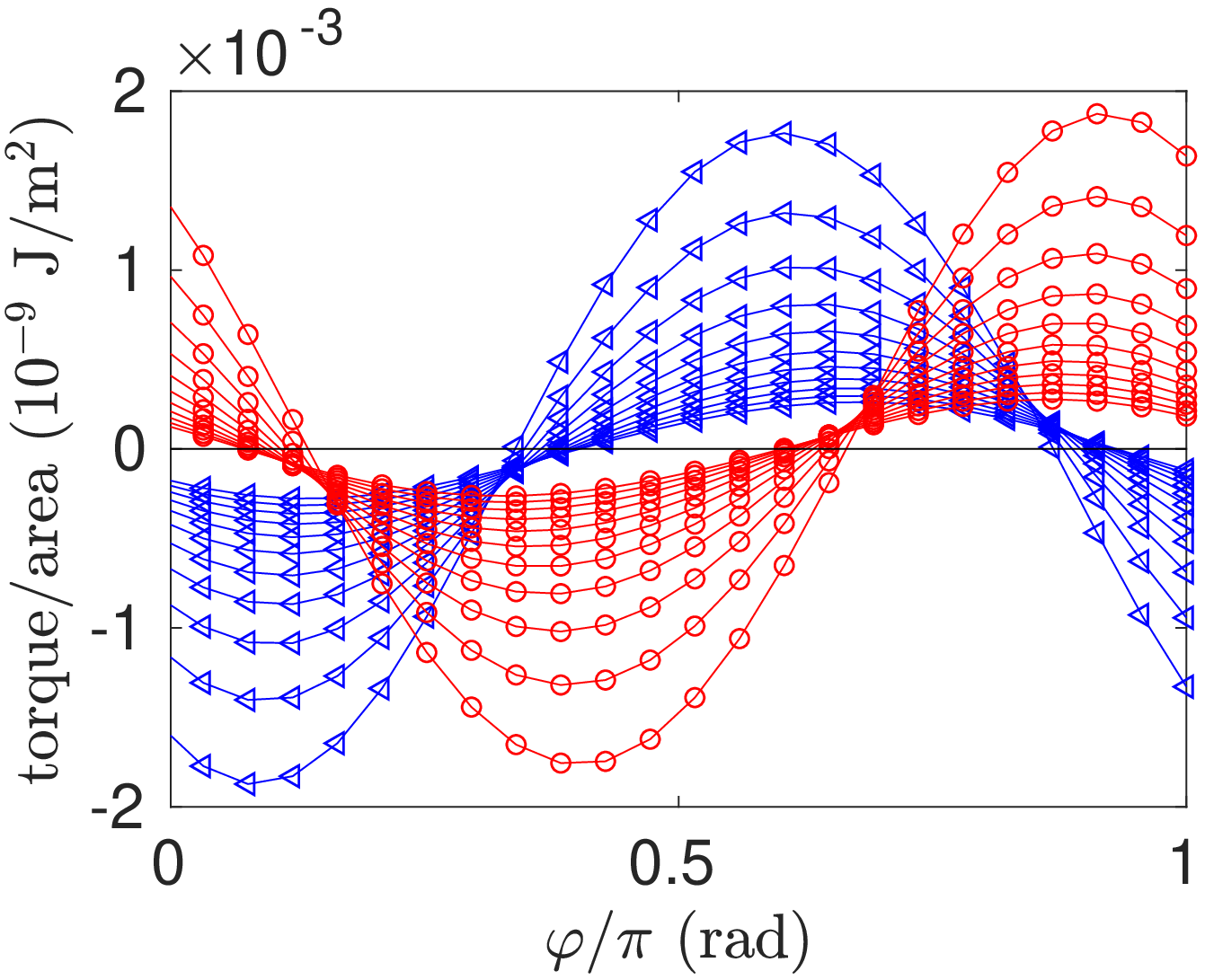}
 \caption{ }
 \label{fig:HomoChiral380-800nm}
 \end{subfigure}
  \caption{The Casimir torque between two cholesteric liquid crystals of identical chirality. (a): homochiral configuration of two left handed crystals (b): homochiral configuration of two right handed crystals. (c):  Casimir torque as a function of misalignment angle at a distance of $a=10$ nm. The different curves are: triangles pointing right: two nematic liquid crystals. Black crosses: unperturbed, case (a). blue squares: perturbative contribution to case (a) . Red circles: unperturbed case (b). orange circles: perturbative  contribution to case (b). (d): The Casimir torque as a function of misalignment angle for distances between 380 an 800 nm. The blue triangles represent case (a) and the red circles represent case (b).}\label{fig:HomoChiral}
\end{figure}

\begin{figure}
    \centering
\begin{subfigure}[t]{0.45\textwidth}
\includegraphics[width=\linewidth]{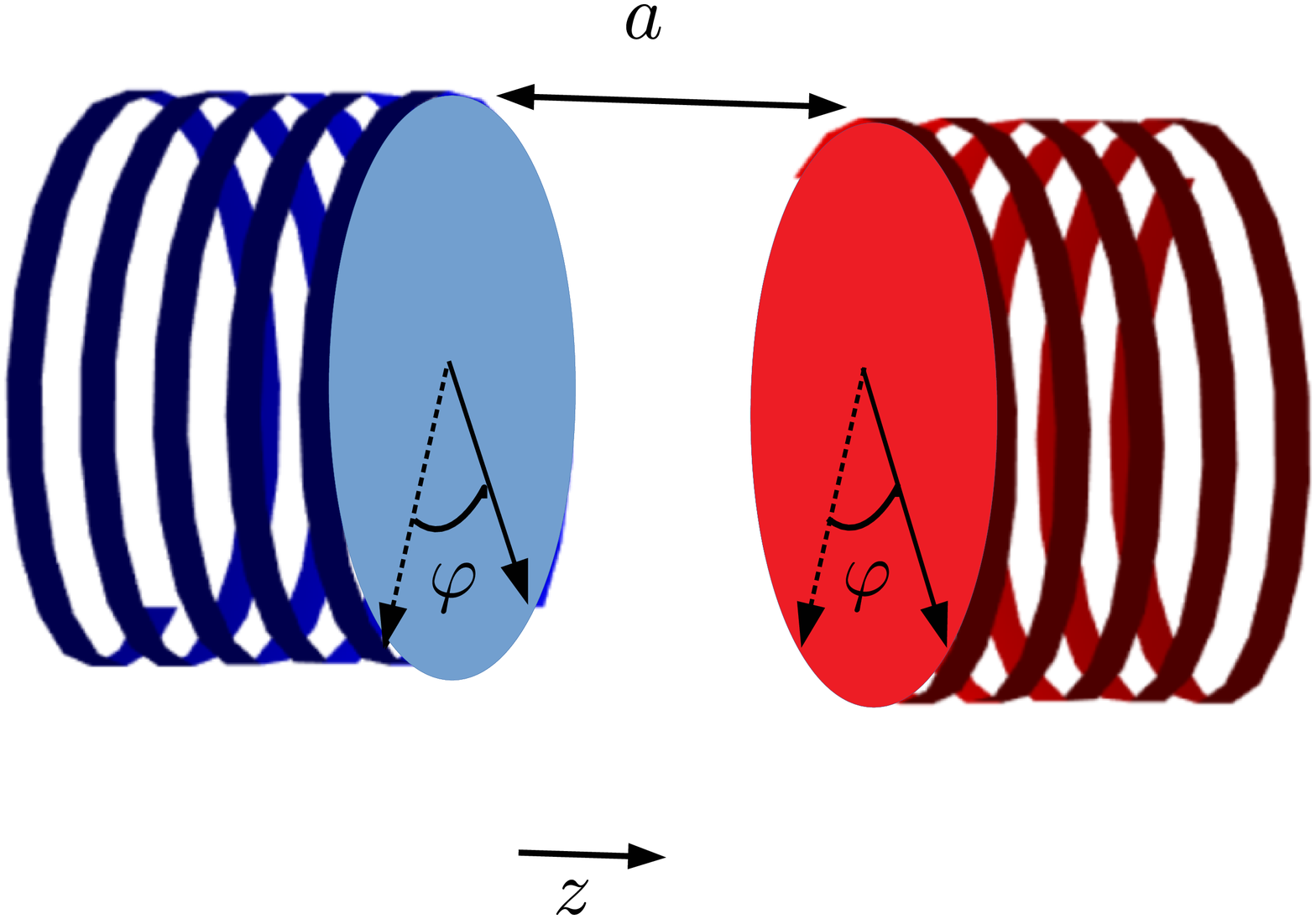}
 \caption{ }
 \label{fig:HeteroChiralLR}
 \end{subfigure}
  \hfill
  \begin{subfigure}[t]{0.45\textwidth}
\includegraphics[width=\linewidth]{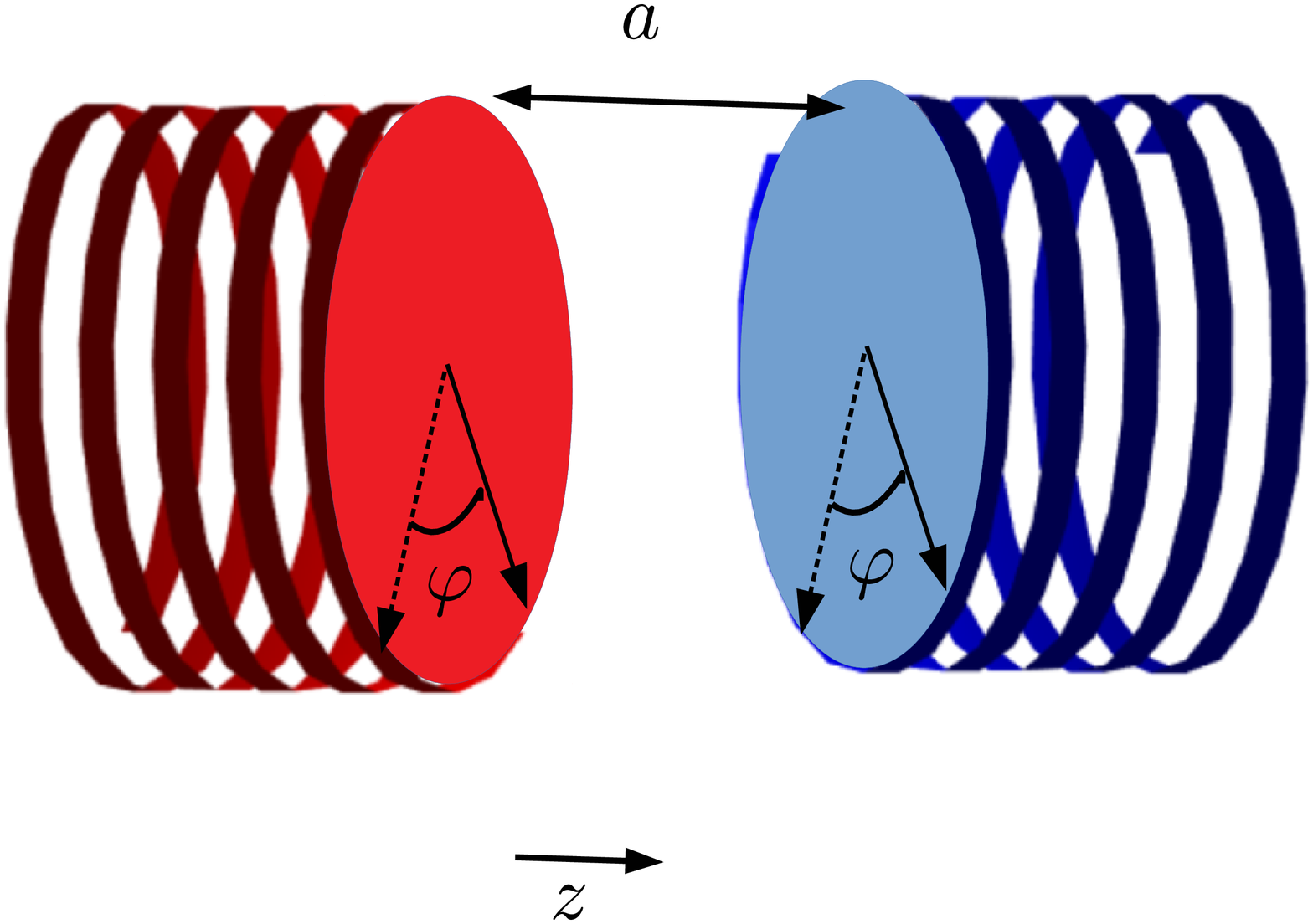}
 \caption{ }
 \label{fig:HeteroChiralRL}
 \end{subfigure}
  \hfill
  \begin{subfigure}[t]{0.45\textwidth}
\includegraphics[width=\linewidth]{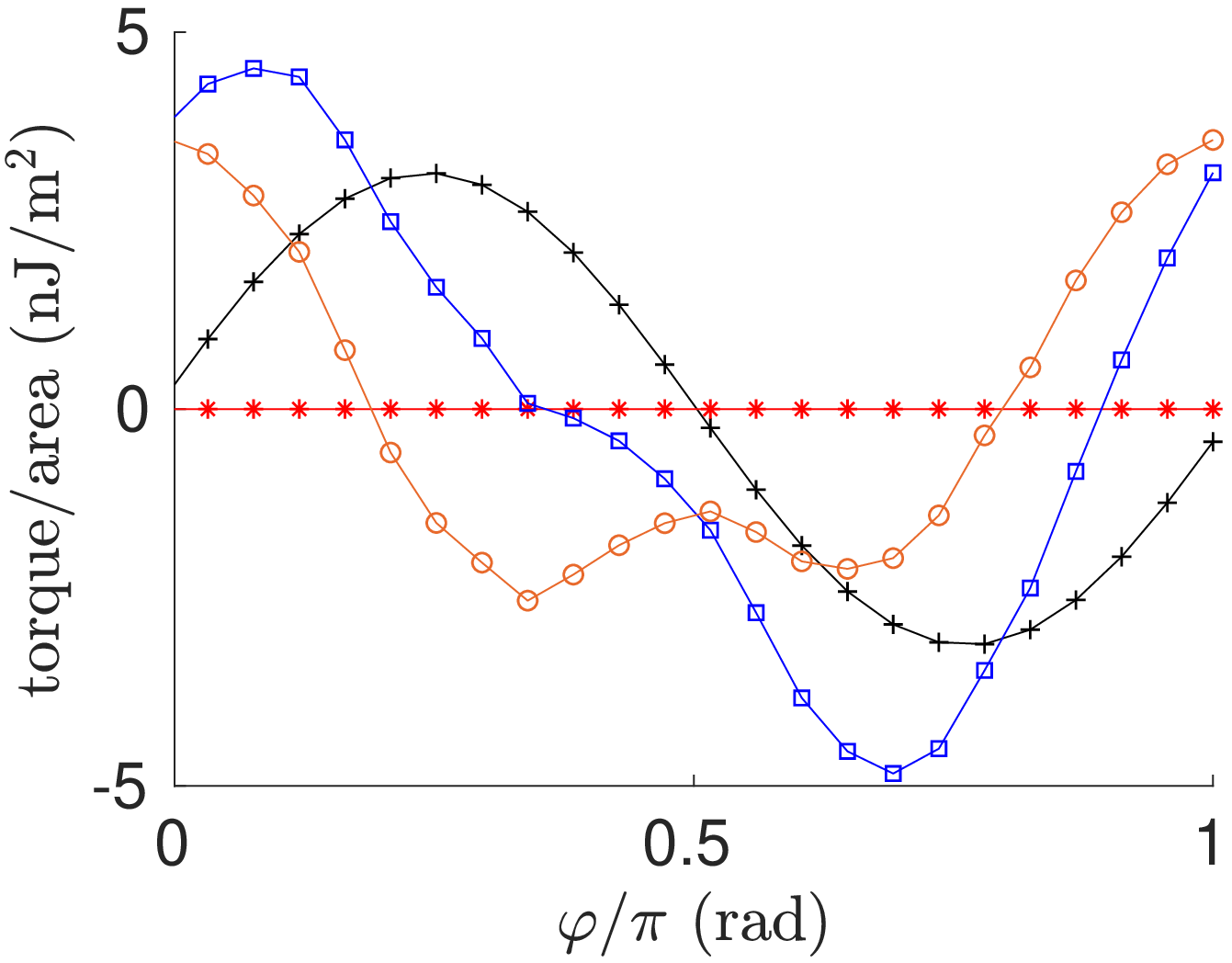}
\caption{ }
\hfill
\label{fig:DeltaTau}
 \end{subfigure}
 \begin{subfigure}[t]{0.45\textwidth}
\includegraphics[width=\linewidth]{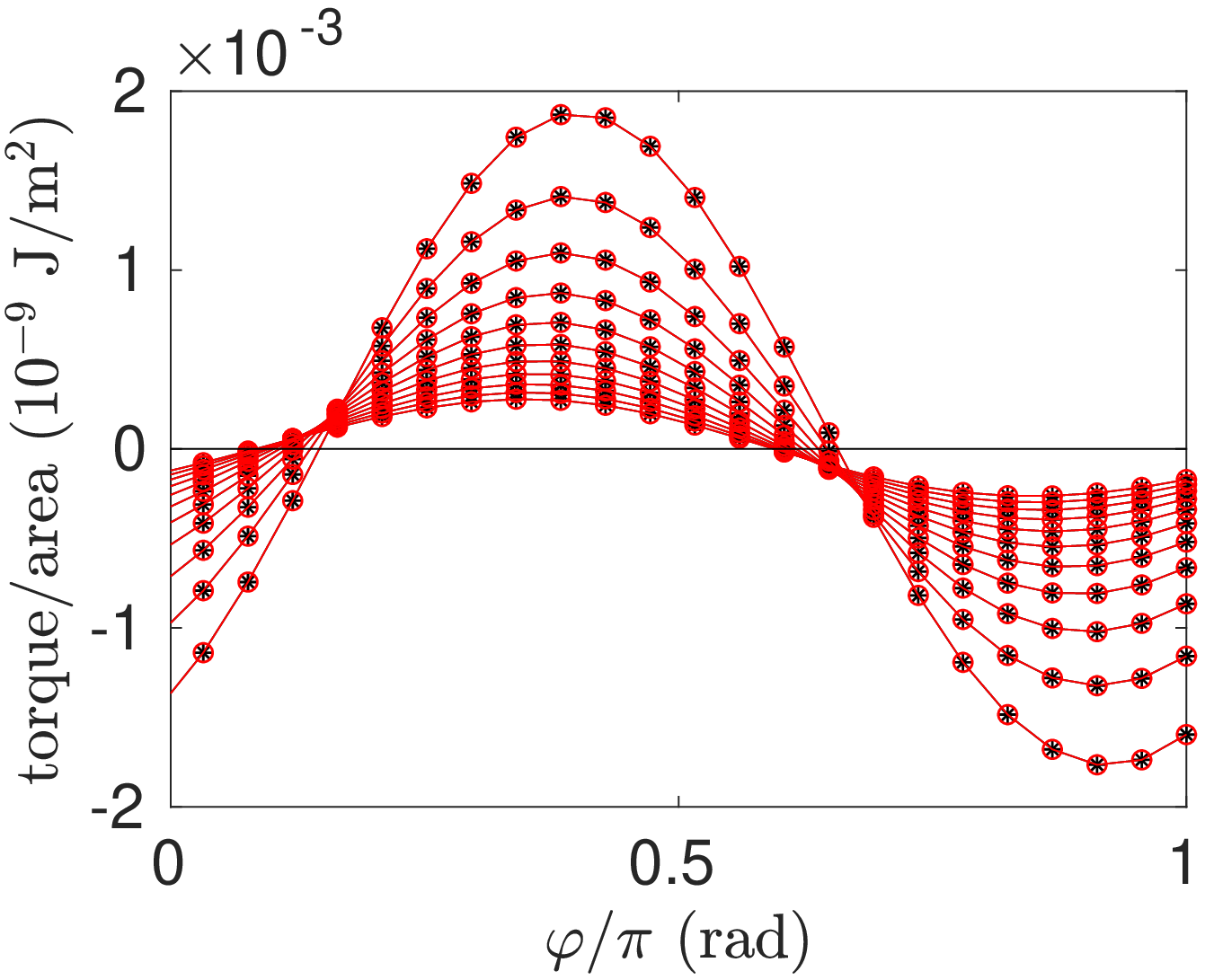}
 \caption{ }
 \label{fig:HeteroChiral380-800nm}
 \end{subfigure}
    \caption{The Casimir torque for the heterochiral case. (a): the left handed crystal is placed on the left, and the right handed crystal is on the right. (b): the reverse of panel (a). (c): several different contributions at 10 nm distance. Black crosses: difference between case (a) and case (b), including only lowest order terms. Red asterisks: idem, including higher order terms. Orange circles: perturbative contribution to case (a), blue squares:  perturbative contribution to case (b). (d): Casimir torque at distances between 380 and 800 nm, both for case (a) (black asterisks), and case (b) (red circles).  }\label{fig:HeteroChiral}
\end{figure}

The results are shown in \cref{fig:BaTiO35CB}. The upper two panels \cref{fig:Left5CBBaTiO3,fig:Right5CBBaTiO3}  display the two cases we distinguish here: the BaTiO$_3$ crystal can be faced with either a right handed or a left handed crystal. \cref{fig:BaTiO35CBphi} shows the Casimir torque as a function of the misalignment angle $\varphi$. For reference, we have included the usual case for a nematic liquid crystal, indicated by the cyan curve with triangles pointing to the right. The leading order term in \cref{eq:final} creates a shift in the angular dependence of the torque of about $\pi/2$ radians, making the proportionality closer to $\cos2\varphi$ than the usual $\sin2\varphi$. The sign of the torque changes depending on the left- or right handedness of the cholesteric liquid crystal, which is a qualitative chirality effect. While the leading order term horizontally shifts the graph, the next order term enhances the amplitude in only one direction, creating a clear deviation from the usual sinusoidal shape of the curve (even though it is still periodic with period $\pi$). For this reason we have verified that 

\begin{equation}\label{eq:energy0}
     \int\limits_0^{\pi}\tau\ud\varphi=0   
\end{equation}
so that no energy is gained or lost during a rotation over a period. Note that the shape of the curve cannot be attributed to retardation effects, which are negligible at a distance of 10 nm. The higher order correction actually is most significant at short distances $\sim$ 10 nm, whereas retardation effects become more pronounced at larger distances. Finally, \cref{fig:BaTiO35CBdistance} shows the distance dependence of the amplitude of the torque. Not only is the amplitude enhanced significantly compared to the nematic case, it also decreases more slowly as a function of distance. Since the configuration of a birefringent and a semi-infinite liquid crystalline sample is presently experimentally accessible \cite{Somers2018}, this could make experimental observation of the torque easier at larger distances. The inset of \cref{fig:BaTiO35CBdistance} shows that the angular dependence remains the same at large distances (between 380 and 800 nm): the positions of the extrema and the nodes do not change as a function of distance. The higher order correction of the BCH formula (\cref{eq:final} is negligible in this separation range,  contributing significantly only at distances of $\sim$ 10 nm.

Next, we proceed to the case of two liquid crystals facing each other. Here we distinguish between the so called homo and heterochiral cases, shown in \cref{fig:HomoChiral,fig:HeteroChiral} respectively. In the former case two right handed or two left handed crystals face each other, and  in the latter case a left  handed crystal is faced with a right handed one.

First we will discuss the homochiral case, as shown in \cref{fig:HomoChiral}. \cref{fig:HomoChiralRight,fig:HomoChiralLeft} illustrate the possible configurations: two right- and two left-handed crystals. \cref{fig:HomoChiralPhi} shows a plot of the Casimir torque as a function of misalignment angle at a distance of 10 nm. For reference, the case of two nematic liquid (light blue triangles) has also been included. As in the previous case, (see \cref{fig:BaTiO35CB}), the leading order term shifts the usual sine to a cosine. However, unlike the previous case, here the higher order terms decrease the amplitude of the torque in one direction, which does give rise to a deviation from the sinusoidal form. This short distance is the only range where the finite pitch length effects, represented by the higher order term in \cref{eq:final}, actually contribute significantly. At this range, it seems as though the torque for the left handed configuration is opposite of that of the right handed one. However, at larger distances it can be seen that this is not the case. In \cref{fig:HomoChiral380-800nm} the Casimir torque is plotted as a function of the misalignment angle for distances between 380 and 800 nm. At this distance range, the higher order term of \cref{eq:final} contribute negligibly. Still even the lower order terms give rise to a significant deviation from the sinusoidal behavior from the previous case. Moreover, the position of the extrema changes with the distance, and the nodes are not on the x-axis. In other words, the angular dependence changes with the separation distance, which is qualitatively different from the previous case or the case of two birefringent materials. The left- and right-handed configurations do not exhibit opposite torques, but the phase difference between the curves changes as a function of distance as well. It is worth pointing out that, if a particular angle is chosen, near one of the nodes, it becomes possible to see the torque change sign as a function of distance. This kind of sign change has also been reported for biaxial materials \cite{Thiyam2018} and Weyl semi-metals \cite{ChenLiang2020}, for example.   

Next we will address the heterochiral case shown in \cref{fig:HeteroChiral}. \cref{fig:HeteroChiralLR,fig:HeteroChiralRL} show the configurations of a left handed crystal on the left and a right handed crystal on the right, and the reverse configuration, respectively. These configurations are expected to be physically indistinguishable. To test this, we plot the different contributions to the torque as a function of misalignment angle again at 10 nm distance, see \cref{fig:DeltaTau}. It can be seen that the difference between the torque of \cref{fig:HeteroChiralLR} and \cref{fig:HeteroChiralRL} is non-negligible if the higher order term in \cref{eq:final} is omitted. (See the black crosses in \cref{fig:DeltaTau}) However, the inclusion of the higher order terms renders this difference negligible. (The red asterisks in \cref{fig:DeltaTau}). So even though the higher order corrections themselves are different (as indicated by the orange circles and blue squares in \cref{fig:DeltaTau}), the total torque is identical for both configurations. Finally, the torque is plotted as a function of the misalignment angle for distances of 380 to 800 nm in \cref{fig:HeteroChiral380-800nm}. Here the torque behaves qualitatively similar to the previous case, with the phase somewhere in between that of the torque for the two homochiral configurations of \cref{fig:HomoChiral380-800nm}. As expected, the torques for the configurations of \cref{fig:HeteroChiralLR,fig:HeteroChiralRL} are identical.

\section{Conclusions and outlook}\label{sec:Conclusions}
In order to understand the effects of cholesteric chirality on the Casimir-Lifshitz torque, we have modeled a cholesteric liquid crystal as a uniaxial planar multilayer system. The system consists of identical layers of equal thicknesses, but the orientation of each layer differs slightly from that of the adjacent one. This model allows us to derive an analytical simplification of the Fresnel matrices as an expansion in this small orientation difference, by means of the Baker-Campbell-Hausdorff (BCH) formula. 

Numerically, we have obtained results that are appreciably different from the case of two semi-infinite birefringent plates, which is a valid approximation for nematic liquid crystals. \cite{Somers2015,Somers2018} In particular, in the case of a birefringent half space facing a cholesteric liquid crystal, the leading order term in the BCH expansion shifts the angular dependence of the torque by about $\pm\pi/2$ radians, depending on the chirality of the liquid crystal. At very short distances of about 10 nm, the higher order BCH term gives rise to a significant deviation from the usual sinusoidal shape of the torque as a function of misalignment angle. Moreover, the amplitude of the torque decreases more slowly as a function of distance than in the nematic case. Since the configuration of a birefringent crystal faced with a liquid crystal is presently experimentally accessible, \cite{Somers2018} we believe that especially this latter result can be useful for detecting the torque at large separation distances.

The case of two cholesteric liquid crystals consists of three physically different configurations: two homochiral ones (two left-handed crystals or two right-handed crystals), and one heterochiral one. In each case, the higher order term of the BCH expansion, which is associated with finite pitch length, contributes significantly at about 10 nm distance.  We have shown that it must be included to obtain physically consistent results in the heterochiral case. At larger distances, the angular dependence of the torque turns out to change as a function of distance, while again significantly deviating from the known sinusoidal behavior. Moreover, for certain values of the misalignment angle, the torque can change sign as a function of distance. Present technology does not yet make it possible to reach a sufficient degree of parallelism to measure the Casimir torque between two liquid crystals with satisfactory accuracy. \cite{Munday2021Review} However, present efforts directed at a resolution of this problem \cite{Sedmik2020} give us hope that it may be possible in the future to measure the torque for such a case.

Possible future endeavors include calculations beyond the approximations used here. For example, one could allow discontinuity in the inhomogeneous dielectric function. This would make it possible to study the effect of a thin water layer between two layers of the cholesteric liquid crystal. Alternatively, one could explore a larger distance range in which case one needs to investigate the effects of a finite total thickness. Finally, the approximate solution of the Maxwell equations obtained here may be of interest for the study of  cholesterics in different contexts, or for the understanding of chiral media in general.

\begin{acknowledgments}
We acknowledge funding from the Key project \#12034019 of the National Natural Science Foundation of China (WB and RP) and the Singapore Ministry of Education Academic Research Fund Tier 1 Grant No. RG160/19(S) (B-SL). 
\end{acknowledgments}

\bibliography{Casimir3}
\end{document}